\documentclass[letter,
  reprint,
  twocolumn,
  superscriptaddress,
  showpacs,
  amsmath,amssymb,
  aps,
  prc,
  floatfix,
]{revtex4-1}

\usepackage{graphicx}
\usepackage{dcolumn}
\usepackage{bm}

\begin{document}

\preprint{APS/123-QED}

\title{Identification of new transitions and mass assignments of levels in $^{143-153}$Pr}

\author{E. H. Wang}
 \affiliation{Department of Physics and Astronomy, Vanderbilt University, Nashville, TN 37235, USA}

\author{A. Lemasson}
 \affiliation{GANIL, CEA/DSM - CNRS/IN2P3, Bd Henri Becquerel, BP 55027, F-14076 Caen Cedex 5, France}

\author{J. H. Hamilton}
\author{A. V. Ramayya}
\author{J. K. Hwang}
\author{J. M. Eldridge}
 \altaffiliation{Present address: Department of Physics, Union University, Jackson, TN 38305, USA}
 \affiliation{Department of Physics and Astronomy, Vanderbilt University, Nashville, TN 37235, USA}

\author{A. Navin}
\author{M. Rejmund}
\affiliation{GANIL, CEA/DSM - CNRS/IN2P3, Bd Henri Becquerel, BP 55027, F-14076 Caen Cedex 5, France}

\author{S. Bhattacharyya}
\affiliation{Variable Energy Cyclotron Centre, 1/AF Bidhan Nagar,
  Kolkata 700064, India}

\author{S. H. Liu}
 \altaffiliation{Present address: Department of Physics and Astronomy, University of Kentucky, Lexington, KY 40506, USA}
 \affiliation{Department of Physics and Astronomy, Vanderbilt University, Nashville, TN 37235, USA}
\author{N. T. Brewer}
 \altaffiliation{Present address: Physics Division, Oak Ridge National Laboratory, Oak Ridge, TN 37831, USA}
 \affiliation{Department of Physics and Astronomy, Vanderbilt University, Nashville, TN 37235, USA}
\author{Y. X. Luo}
 \affiliation{Department of Physics and Astronomy, Vanderbilt University, Nashville, TN 37235, USA}
\author{J. O. Rasmussen}
 \affiliation{Lawrence Berkeley National Laboratory, Berkeley, CA 94720, USA}
\author{H. L. Liu}
 \affiliation{Department of Physics, Xi`an Jiaotong University, Xi'an 710049, People's Republic of China}
 \author{H. Zhou}
 \affiliation{Department of Physics, Peking University, Beijing 100871, People's Republic of China}
\author{Y. X. Liu}
 \affiliation{Department of Physics, Huzhou University, Huzhou 313000, People's Republic of China}
\author{H. J. Li}
 \affiliation{Department of Physics, Tsinghua University, Beijing 100084, People's Republic of China}
\author{Y. Sun}
 \affiliation{Department of Physics, Shanghai Jiao Tong University, Shanghai 200240, China}
\author{F. R. Xu}
 \affiliation{Department of Physics, Peking University, Beijing 100871, People's Republic of China}
\author{S. J. Zhu}
 \affiliation{Department of Physics, Tsinghua University, Beijing 100084, People's Republic of China}
\author{G. M. Ter-Akopian}
\author{Yu. Ts. Oganessian}
 \affiliation{Joint Institute for Nuclear Research, RU-141980 Dubna, Russian Federation}

\author{M. Caama\~{n}o}
\affiliation{USC, Universidad de Santiago de Compostela, E-15706
Santiago de Compostela, Spain}

\author{E. Cl\'ement}
\author{O. Delaune}
\author{F. Farget}
\author{G. de France}
\author{B. Jacquot}

\affiliation{GANIL, CEA/DSM - CNRS/IN2P3, Bd Henri Becquerel, BP 55027, F-14076 Caen Cedex 5, France}

\date{\today}

\begin{abstract}
\noindent $\bf{Background:}$  The previously reported  levels assigned
to  $^{151,152,153}$Pr   have  recently  been   called  into  question
regarding their mass assignment.\\
$\bf{Purpose:}$  Clarify  the above  questioned  level assignments  by
measuring  $\gamma$-transitions tagged  with  A and  Z  in an  in-beam
experiment in addition to the measurements from $^{252}$Cf spontaneous
fission (SF) and establish  new spectroscopic information from $N=84$ to
$N=94$ in the Pr isotopic chain.\\
$\bf{Methods:}$  The isotopic  chain $^{143-153}$Pr  has  been studied
from the  spontaneous fission of  $^{252}$Cf by using  Gammasphere and
also from  the measurement of the prompt  $\gamma$-rays in coincidence
with  isotopically-identified  fission  fragments  using  VAMOS++  and
EXOGAM at  GANIL. The latter were  produced using $^{238}$U  beams on a
$^{9}$Be  target   at  energies   around  the  Coulomb   barrier.  The
$\gamma$-$\gamma$-$\gamma$-$\gamma$  data  from  $^{252}$Cf  (SF)  and
those from the  GANIL in-beam A- and Z-gated  spectra were combined to
unambiguously   assign   the  various   transitions   and  levels   in
$^{151,152,153}$Pr and other isotopes.\\
$\bf{Results:}$ A band  of 3 new transitions added  to the known level
in $^{145}$Pr, 9 new transitions in two new bands in $^{147}$Pr, 6 new
transitions in a  new level scheme for $^{148}$Pr,  two new bands with
17  new  transitions  in  $^{149}$Pr  and  2 new  bands  with  11  new
transitions    in     $^{150}$Pr    were    identified     by    using
$\gamma$-$\gamma$-$\gamma$   and   $\gamma$-$\gamma$-$\gamma$-$\gamma$
coincidences  and  A  and  Z  gated  $\gamma$-$\gamma$  spectra.   The
transitions and levels previously assigned to $^{151,153}$Pr have been
confirmed by the  (A,Z) gated spectra.  Small changes  have been made to
their original  level schemes. The transitions  previously assigned to
$^{152}$Pr are  now assigned to $^{151}$Pr  on the basis  of the (A,Z)
gated spectra. Two new bands with 20 new transitions in $^{152}$Pr and
one new band with 7  new transitions in $^{153}$Pr are identified from
the  $\gamma$-$\gamma$-$\gamma$-$\gamma$ coincidence  spectra  and the
(A,Z) gated spectrum. In addition, new $\gamma$-rays are also reported
in $^{143-146}$Pr.\\
$\bf{Conclusions:}$  New   levels  of  $^{145,147-153}$Pr   have  been
established,   reliable   mass    assignments   of   the   levels   in
$^{151,152,153}$Pr  have  been  given  in  the present  work  and  new
transitions  have been  identified in  $^{143-146}$Pr showing  the new
avenues that are opened by combining the two experimental approaches.
\end{abstract}

\pacs{23.20.Lv, 25.85.Ca, 21.10.-k, 27.70.+q}
\keywords{Suggested keywords}
\maketitle


\section{\label{sec:level1}introduction}

Studies  of nuclear  energy levels  over long  isotopic  chains reveal
structural changes  as a function of  N and provide  important test of
nuclear models. Spontaneous fission  (SF) has provided a good approach
to study  nuclei over long  isotopic chains \cite{Ham95}.  In  SF, the
new  transitions in  a  certain isotope  are  generally identified  by
gating on the known transitions  in the particular isotope observed in
$\beta$-decay or  by gating on known transitions  in the complementary
fission fragment, which are usually less neutron-rich and well studied
in  most  of  the  cases.   Usually,  this  procedure  gives  reliable
identifications. However, in some cases, when the $\gamma$-spectrum is
complex, the overlapping of  transition energies in different isotopes
could lead to a wrong mass identification of the bands.

Recently prompt  $\gamma$ ray spectroscopy of  fully identified (A,~Z)
fission  fragments  produced  in fusion-fission  and  transfer-fission
reactions  around  the  Coulomb  barrier  \cite{Nav,Nav14}  have  been
reported where  due to the advantage of  unambiguously identifying the
fragments, the  assignments of $\gamma$-rays  to a particular A  and Z
are directly  obtained. The combination  of the traditional  high fold
gamma-coincidence method \cite{Ham95} and this method \cite{Nav,Nav14}
is  expected  to  strengthen  explorations and  understanding  of  the
evolution of nuclear structure as a function of both isospin and spin.

Historically, levels of several  Pr isotopes have been identified from
$^{252}$Cf   and   $^{248}$Cm   SF  \cite{Hwa00,Hwa10,Rza,Liu}.   Only
$^{151,153}$Pr  were reported to  have possible  octupole correlations
between   parity-doublet   bands   \cite{Hwa10}.   As   mentioned   in
Ref.~\cite{Hwa10}  the  octupole   correlations  in  this  region  are
associated with $\Delta$N~=~1, $\Delta$j~=~3 and $\Delta l$~=~3 orbital
pairs such  as $\pi  d_{5/2}- h_{11/2}$ near  Z =56 and  $\nu f_{7/2}-
i_{13/2}$ near N =88. The  mass assignments of the previously reported
levels assigned  to $^{151,152,153}$Pr \cite{Hwa10,Liu}  have recently
been called into question \cite{Mal}. The $\gamma$-ray transitions and
levels in  $^{151,153}$Pr reported  in the previous  work \cite{Hwa10}
were assigned  to $^{152,154}$Pr respectively  in Ref.~\cite{Mal}. The
two  bands  proposed  to  be  in $^{152}$Pr  in  Ref.~\cite{Liu}  were
assigned to  $^{151,153}$Pr separately \cite{Mal}.   The assignment of
gamma transitions to their  corresponding nuclei is challenging due to
the closely spaced  transitions and the complexity of  the spectra. In
order to  have an  unambiguous identification of  the nuclei in  the A
$\sim$ 150 mass region, two  different techniques have been combined in
the present work to investigate  the high spin states and the possible
octupole correlations for neutron rich Pr isotopes.  In this paper, we
provide  new analysis  of  both $^{252}$Cf  SF and  $^{238}$U+$^{9}$Be
induced  fission data \cite{Nav,Nav14}  with direct  identification of
fission fragment mass and Z to give reliable assignments of the levels
and  transitions   in  these  Pr   isotopes.  New  level   schemes  of
$^{145,147-153}$Pr and new  transitions in $^{143-146}$Pr are reported
in the  present work.  The possibility of  the occurrence  of octupole
correlations in  the band  structures of $^{149,151}$Pr  are indicated.

\section{Experimental method}

Two  complementary methods  have been  used to  investigate  the level
structure  of   Pr  isotopes,  which  include   both  the  unambiguous
identification  of the  mass  (A) and  the  proton number  (Z) of  the
emitting fission  fragment using  a large acceptance  spectrometer for
in-beam measurements  and the high fold data  from spontaneous fission
of a $^{252}$Cf source. These complementary methods have allowed us to
identify  new  transitions and  extend  the  level  schemes to  higher
spins. In the present work the new transitions identified using (A,~Z)
gated  'singles'  prompt  $\gamma$-ray  spectroscopy did  not  require
knowledge  of  the  spectroscopic  information  of  the  complementary
fragment.   This  allowed  the  study  of very  neutron  rich  nuclei,
combining  the  unique  in-beam  identification  with  the  high  fold
$\gamma-\gamma-\gamma-\gamma$ coincidences from  SF. In the following,
double  gated  coincidence  spectra  have a  variable  energy  binning
ranging  from 0.7~keV/channel  at 100~keV  to 1~keV/channel  at 1~MeV.
The  triple  gated  coincidence   spectra  have  a  fixed  binning  of
1.3~keV/channel.

\subsection{$^{238}$U + $^{9}$Be induced fission}

The measurements of transfer and fusion induced fission were performed
at GANIL using a $^{238}$U beam at 6.2 MeV/u, with a typical intensity
of  0.2 pnA,  impinging on  a  10-$\mu$m thick  $^{9}$Be target.   The
advantage of the inverse kinematics  used in this work is that fission
fragments are forward focussed and have a large velocity, resulting in
both  an  efficient  detection  and  isotopic  identification  in  the
spectrometer. A single magnetic  field setting of the large-acceptance
spectrometer VAMOS++ \cite{Rej11}, possessing a momentum acceptance of
around $\pm$ 20\%, placed at 20$^\circ$ with respect to the beam axis,
was used  to identify uniquely  the fission fragments.   The detection
system (1$\times$0.15 m$^{2}$ ) at the focal plane of the spectrometer
was  composed of  (i) a  Multi-Wire Parallel  Plate  Avalanche Counter
(MWPPAC), (ii) two Drift  Chambers (x,y), (iii) a Segmented Ionization
Chamber ($\Delta$E), and (iv) 40  silicon detectors arranged in a wall
structure (E$_r$).   The time of  flight (TOF) was obtained  using the
signals from  the two  MWPPACs, one located  after the target  and the
other at  the focal plane  (flight path $\sim$7.5 m).   The parameters
measured at the focal plane  [(x,y), $\Delta$E, E$_r$, TOF] along with
the known magnetic field were  used to determine, on an event-by-event
basis, the mass  number (A), charge state (q),  atomic number (Z), and
velocity vector after the reaction for the detected fragment. Isotopic
identifications  of  elements were  made  up  to  $Z=63$ with  a  mass
resolution of $\Delta A/A\sim 0.4$\% \cite{Nav13}. The prompt $\gamma$
rays  were measured  in coincidence  with  the isotopically-identified
fragments,  using  the EXOGAM  array  \cite{ExoGaM}  consisting of  11
Compton-suppressed  segmented clover  HpGe detectors  (15 cm  from the
target).  The  velocity of  the fragment along  with the angle  of the
segment  of the  relevant  clover  detector were  used  to obtain  the
$\gamma$-ray energy in the rest frame of the emitting fragment.  Error
on  the $\gamma$-ray energies  of the  strong transitions  is 0.5~keV,
while  for the  weak transitions  it could  be as  much as  1~keV.  As
compared to  the results  presented in Ref.   \cite{Nav13} for  the Zr
isotopes the present work is the result of further improvements in the
analysis especially improving the Z identification and also involves a
larger  data  set.   Figure~\ref{fig:fig2D}  shows a  two  dimensional
spectra of A vs 'singles'  E$_\gamma$ for the Pr isotopes using such
a data set which also allows one to view directly the evolution of the
various transitions as a function of mass number.

\subsection{$^{252}$Cf spontaneous fission}

\begin{figure}[t]
\includegraphics[width=\columnwidth]{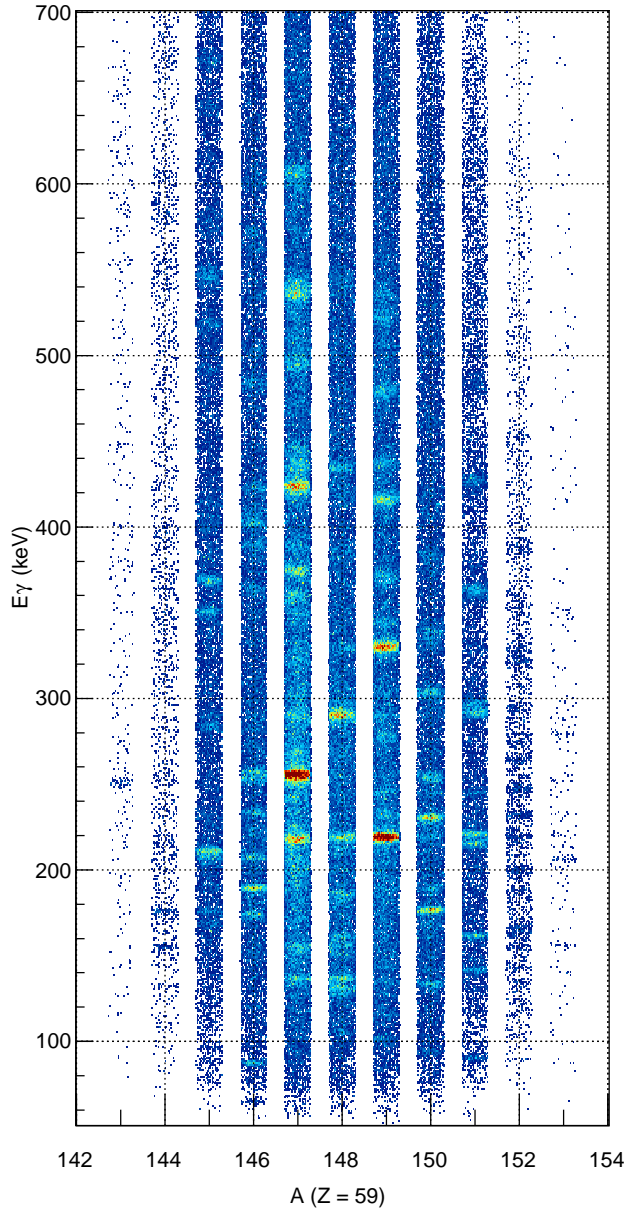}
\caption{\label{fig:fig2D}
(color  online)
Doppler-corrected $\gamma$-ray energy
as a function of the mass number (A) of the Pr (Z=59)  fragment identified in VAMOS${++}$. }
\end{figure}
\begin{figure}
\centering\includegraphics[width=\columnwidth]{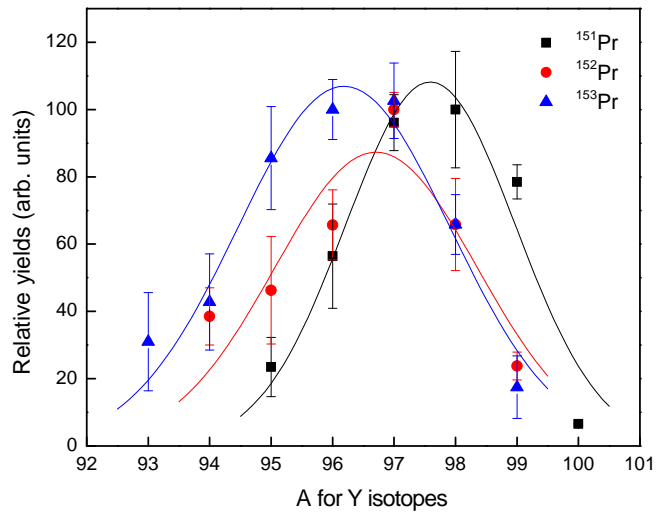}
\caption{\label{yield}(Color online) Relative  yield curves of yttrium
  by  gating  on  transitions  in $^{151}$Pr  (black  square  symbol),
  $^{152}$Pr  (red circular  symbol) and  $^{153}$Pr  (blue triangular
  symbol)  from  $^{252}$Cf  SF  data.  Each Pr  gate  yield  set  are
  normalized to the highest  Y yield for comparison.  Fission partners
  of  3n  channels  for  $^{151}$Pr,  $^{152}$Pr  and  $^{153}$Pr  are
  $^{98}$Y, $^{97}$Y and $^{96}$Y respectively.  The gaussian fittings
  for Pr gate yields are also shown as lines in the same colors.}
\end{figure}
\begin{figure}[h]
 \includegraphics[width=\columnwidth]{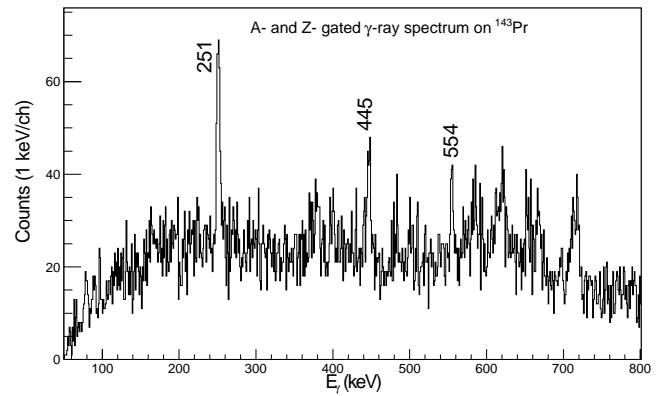}
 \caption{\label{143mass}Partial   $^{143}$Pr    A-   and   Z-   gated
   $\gamma$-ray  spectrum obtained from  $^{238}$U +  $^{9}$Be induced
   fission data.}
\end{figure}

The  experiment  with  $^{252}$Cf  was  carried out  at  the  Lawrence
Berkeley National Laboratory(LBNL). A 62 $\mu$Ci $^{252}$Cf source was
sandwiched between  two Fe foils of thickness  10mg/cm$^{2}$. By using
101  Ge   detectors  of  Gammasphere,   the  data  were   sorted  into
5.7$\times$10$^{11}$   $\gamma$-$\gamma$-$\gamma$   and  higher   fold
$\gamma$           events           and           1.9$\times$10$^{11}$
$\gamma$-$\gamma$-$\gamma$-$\gamma$    and   higher    fold   $\gamma$
coincident events. These $\gamma$ coincident data were analyzed by the
RADWARE software package \cite{Rad}.  Gamma-ray energies of the strong
transitions  have errors  of  0.1~keV  while  the errors  on the  weak
transitions could be as much as 0.5~keV.

Another similar experiment  was performed earlier at LBNL  by using 72
Ge detectors of Gammasphere with  a 28 $\mu$Ci $^{252}$Cf source. Less
$\gamma$-$\gamma$-$\gamma$  and higher  fold  coincidence events  were
collected  in this  experiment but  the data  were built  according to
several discrete coincidence time windows ranging from 4ns to 500ns of
the $\gamma$-rays \cite{Hwa98,Hwa03}.

To   independently  confirm  the   mass  assignments   obtained  above
(Fig.~\ref{fig:fig2D}),  the relative  yield curves  of  the yttrium
partner    isotopes    coincident    with   $^{151,152,153}$Pr    were
measured.  Fig.~\ref{yield} is  a set  of Y  yield curves  measured by
double  or  triple  gating   on  transitions  in  $^{151-153}$Pr.  The
intensity summations of all  the observable transitions which directly
feed the ground state were  used as a representative of the respective
Y yields. The $^{97}$Y isotope has an isomeric state at a level energy
of  667.5~keV~(1.17s).  Therefore,  the  intensity  summation  of  two
transitions (668.6 and 989.9~keV), which feed the 667.5~keV state, was
used instead. In the $^{252}$Cf  binary spontaneous fission, a pair of
correlated partners is produced and followed by neutron emission after
fission. The yield is generally maximized  at the 3n or 4n channel \cite{Wah} In
Fig.~\ref{yield}, the  highest yields were populated  at around $^{98,
  97, 96}$Y  for $^{151, 152,  153}$Pr respectively, which are  all 3n
reaction  channels for  these  Pr isotopes.  Further Gaussian  fitting
analysis  for  the  curves   indicates  that  the  fitting  peaks  for
$^{151,152,153}$Pr  are located at  the respective  3.4, 3.3  and 2.8n
channels. These  yield distribution curves  of the Y  partner isotopes
are   consistent   with   the    mass   assignments   of   levels   in
$^{151,152,153}$Pr.

\section{Experimental results}

In this section new  transitions for $^{143-147}$Pr are reported using
in-beam fission  whereas the  new level schemes  of $^{145,147-153}$Pr
were obtained by combining both the data sets.

\subsection{$^{143}$Pr}
The A-  and Z- gated $\gamma$-ray  spectrum on $^{143}$Pr  is shown in
Fig.~\ref{143mass}.   Three new  transitions 251(1),445(1)  and 554(1)
keV were identified. Other transitions are not labelled.

\subsection{$^{144}$Pr}
\begin{figure}
 \includegraphics[width=\columnwidth]{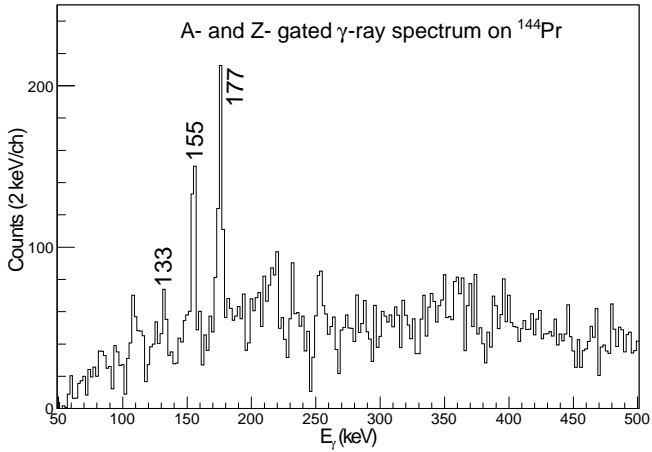}
 \caption{\label{144mass}Partial   $^{144}$Pr  mass-   and   Z-  gated
   $\gamma$-ray  spectrum obtained from  $^{238}$U +  $^{9}$Be induced
   fission data.}
\end{figure}

The A- and Z- gated $\gamma$-ray spectrum on $^{144}$Pr is shown in
Fig.~\ref{144mass}.   The  133(1)~keV  peak  could  be  the 133.5~keV
transition  identified  in  $^{144}$Ce $\beta$-decay  \cite{Dal}.   It
should be pointed  out that the unbiased 'singles'  nature of the data
allows the identification of low multiplicity transitions seen in beta
decay.  In  addition two  new transitions 155(1)  and 177(1)~keV were
also identified.   Due to the complexity  of the spectrum  and lack of
$\gamma-\gamma$   coincidences   in   this  odd-odd   nucleus,   other
transitions  above 200~keV  are not  labelled and  no level  scheme is
presented.

\subsection{$^{145}$Pr}

\begin{figure}
  \includegraphics[width=\columnwidth]{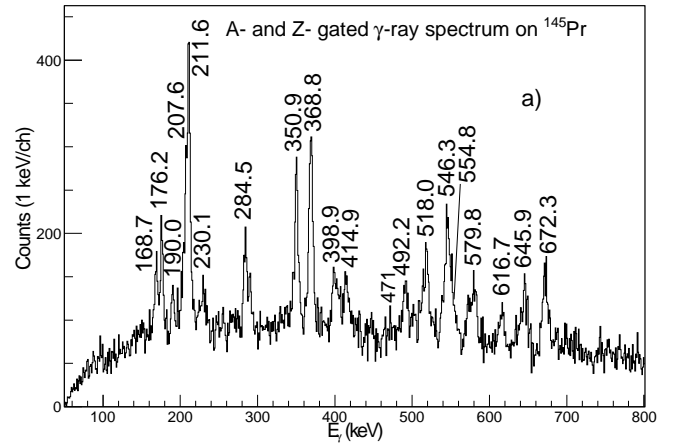}
  \includegraphics[width=\columnwidth]{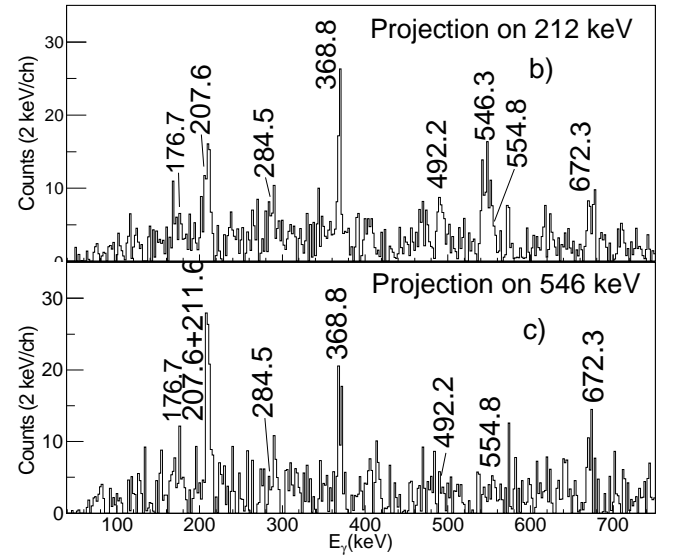}
  \caption{\label{145mass}Partial   $^{145}$Pr  mass-  and   Z-  gated
    $\gamma$-ray  spectra obtained from  $^{238}$U +  $^{9}$Be induced
    fission data.  Part (a)  is 'single' $\gamma$-ray spectrum. Part (b)
    is a gated spectrum on the 211.6~keV transition and part (c) gated
    on the 546~keV transition, respectively.}
\end{figure}

\begin{figure}[t]
 \includegraphics[width=0.7\columnwidth]{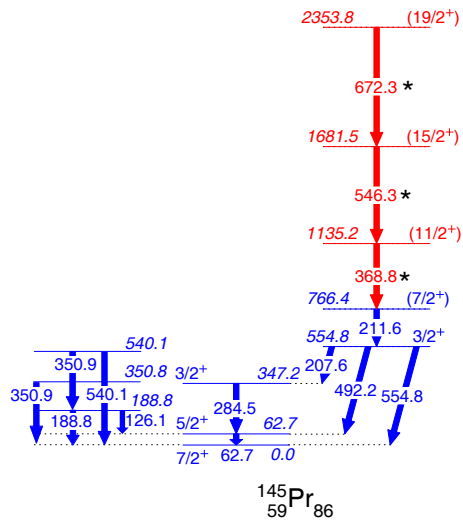}
 \caption{\label{145level}(Color  online)  The  new  level  scheme  of
   $^{145}$Pr in  the present work. Transitions  and levels previously
   reported in $\beta$ decay work are labeled in blue. New ones in the
   current work are  labeled in red and with an asterisk.  The width  of the transitions in
   the level scheme in the current work does not indicate the relative
   intensity.}
\end{figure}
\begin{figure}[ht]
  \begin{minipage}[t]{\columnwidth}
  \includegraphics[width=0.9\columnwidth]{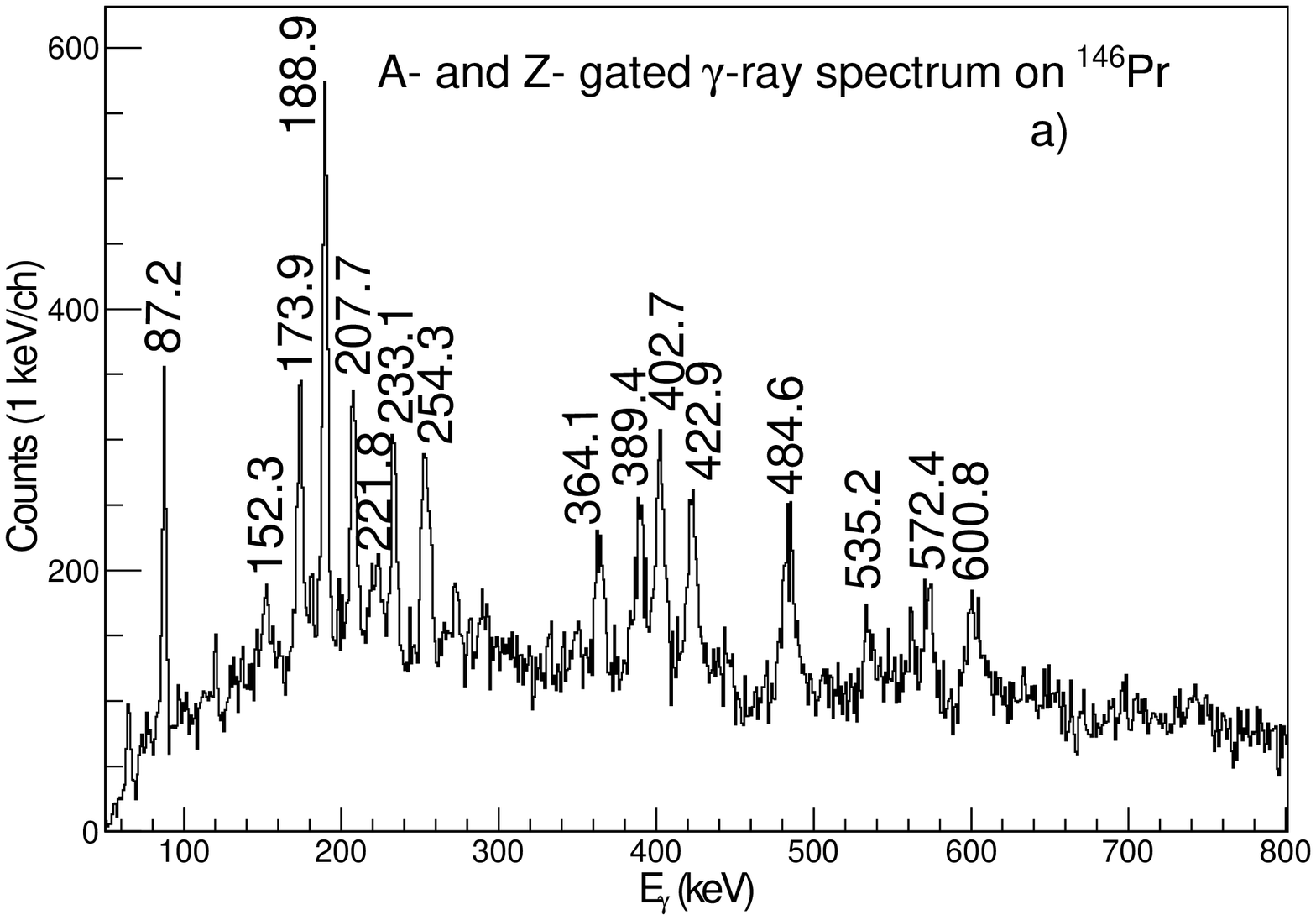}
  \end{minipage}
  \begin{minipage}[t]{\columnwidth}
  \includegraphics[width=0.9\columnwidth]{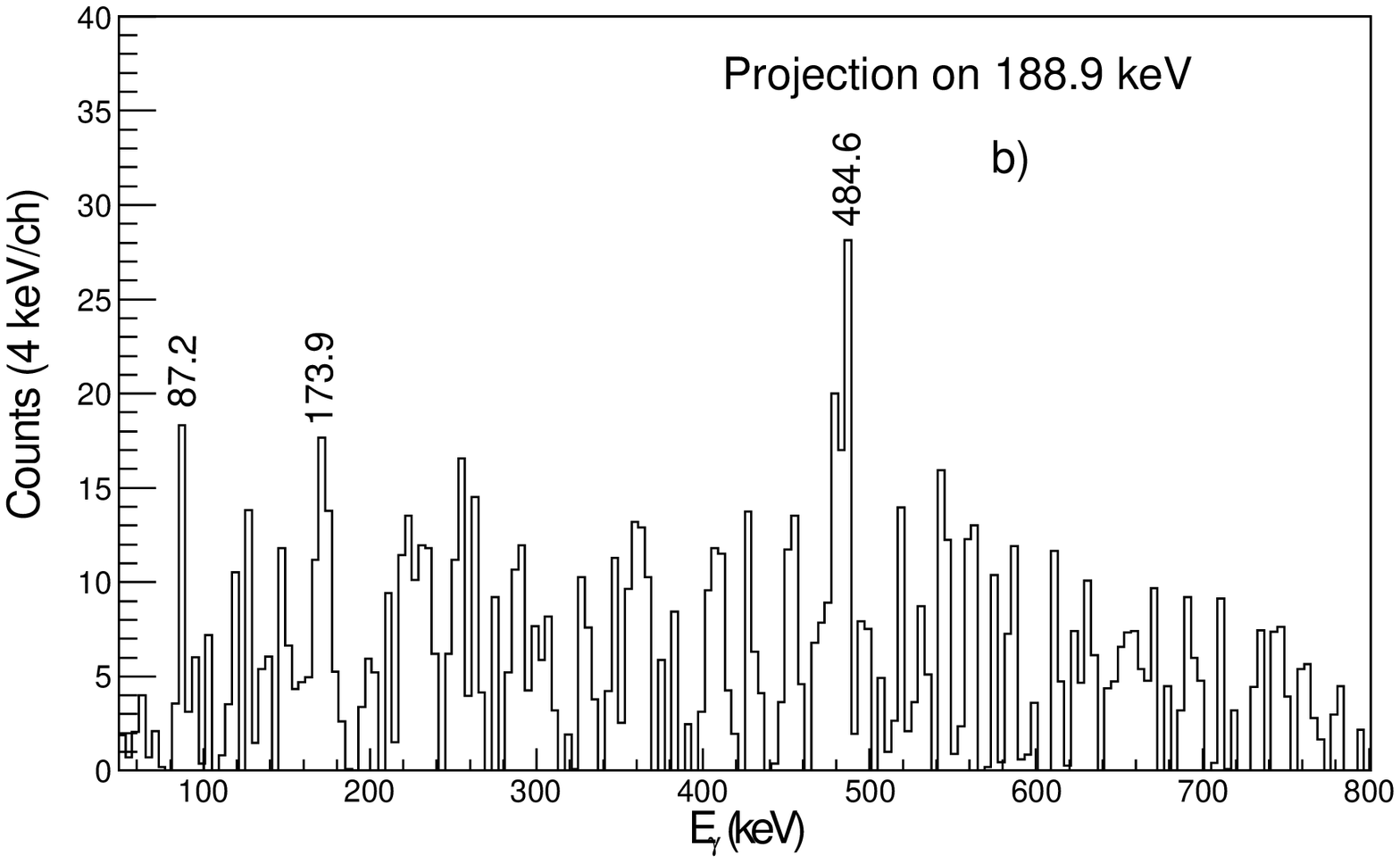}
  \caption{\label{146mass}Partial   $^{146}$Pr  A-  and   Z-  gated
    $\gamma$-ray  spectra obtained from  $^{238}$U +  $^{9}$Be induced
    fission data.  Part (a)  is 'single' $\gamma$-ray spectrum. Part (b)
    is a spectrum gated on 188.9~keV transition.}
 \end{minipage}
\end{figure}

The A- and Z- gated $\gamma$-ray spectrum on $^{145}$Pr is shown in
Fig.~\ref{145mass}(a).  The  207.6, 211.6,  284.5 and 492.2~keV peaks
were previously identified in $^{145}$Ce $\beta$-decay \cite{Bau}. The
350.9~keV  peak could be an  overlap of a 350.9~keV transition (decay
from  350.9~keV  level  to the  ground  state seen in $\beta$-decay) and  another 350.9~keV
transition (decay from 540.1~keV  level to a 188.8~keV level in $\beta$-decay) identified
in  $\beta$-decay  \cite{Bau}.   Fig.~\ref{145mass}(b) and  (c)  shows
$\gamma$-ray spectra  gated on the 211.6~keV  and 546~keV transitions
respectively. In  these spectra,  the coincident 204.0,  207.6, 368.8,
492.2, 546.3 and  672.3~keV transitions can be  seen.  From the energy
spacing of the  211.6, 368.8, 546.3 and 672.3~keV transitions and the
intensities  shown in  Fig.~\ref{145mass}(a), these  $\gamma$-rays are
possibly E2 transitions in a rotational band. Thus, spins and parities
of levels in this band  are tentatively assigned. The level scheme for
$^{145}$Pr is  shown in Fig.~\ref{145level}. It should  be pointed out
that in the  beta decay measurements the 211.6 is  weak whereas in the
present work  it is very  intense, showing the complementarity  of the
in-beam and decay  work.  There is coincidence evidence  in the (A,~Z)
gated data for a 350-415 keV cascade feeding a 168~keV transition.  It
is not  definitive that this cascade feeds  into the 207-212~cascades,
so it  is not placed in the  level scheme.  The (A,~Z) gated $\gamma$
coincidence   data   (Fig.~\ref{145mass})   indicate   the   176.7~keV
transition is in coincidence with the 207-212-284-368-546~keV cascade,
but it is not clear where to place it. The observed 190.0, 230.1, 518.0, 579.8, 616.7 keV transitions are also new but we have no coincidence data to place them.
\subsection{$^{146}$Pr}

The A- and Z- gated $\gamma$-ray spectrum on $^{146}$Pr is shown in
Fig.~\ref{146mass}(a). The 87.2~keV  transition can be identified with
that     previously      known     in     $^{146}$Ce     $\beta$-decay
\cite{Yam80}. Fig.~\ref{146mass}(b)  is a spectrum gated  on the 188.9
keV transition. In this spectrum, the coincident 87.2, 173.9 and 484.6
keV transitions can be seen.  With only these data it was not possible
to build a level scheme.

\subsection{$^{147}$Pr}

\begin{figure}[]
 \includegraphics[width=\columnwidth]{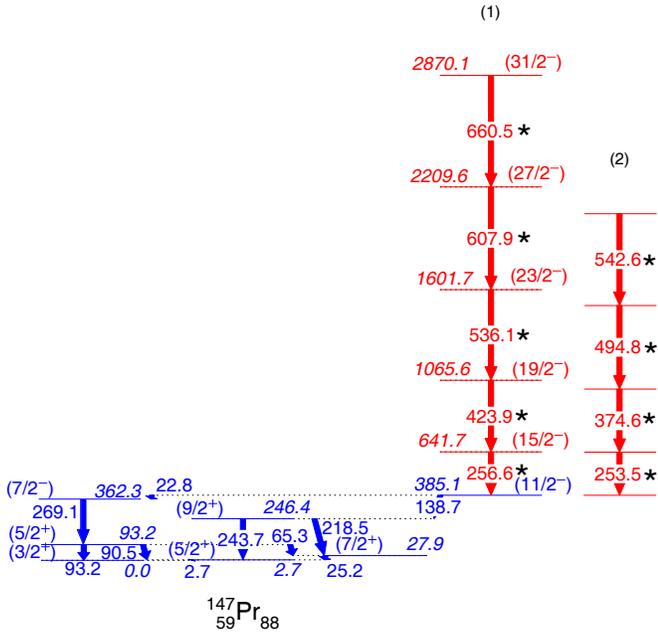}
 \caption{\label{147level}(Color  online)  The  new  level  scheme  of
   $^{147}$Pr in  the present work. Transitions  and levels previously
   reported in $\beta$ decay work are labeled in blue. New ones in the
   current work are labeled in red and with an asterisk. It is possible there is a low energy transition between the 256.6 and 138.7 keV transitions as discussed in the text.}
\end{figure}
\begin{figure}
 \includegraphics[width=\columnwidth]{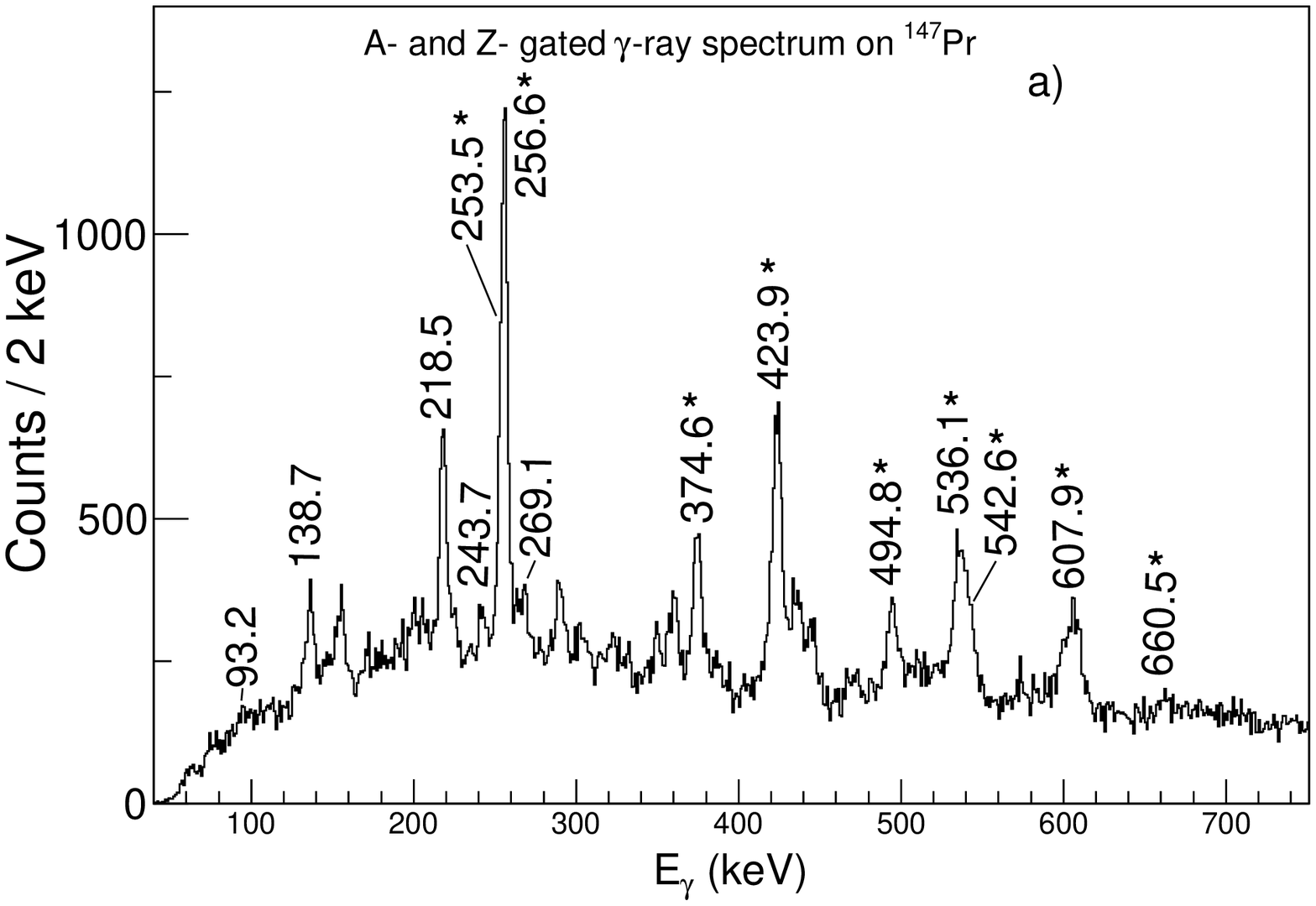}
 \includegraphics[width=\columnwidth]{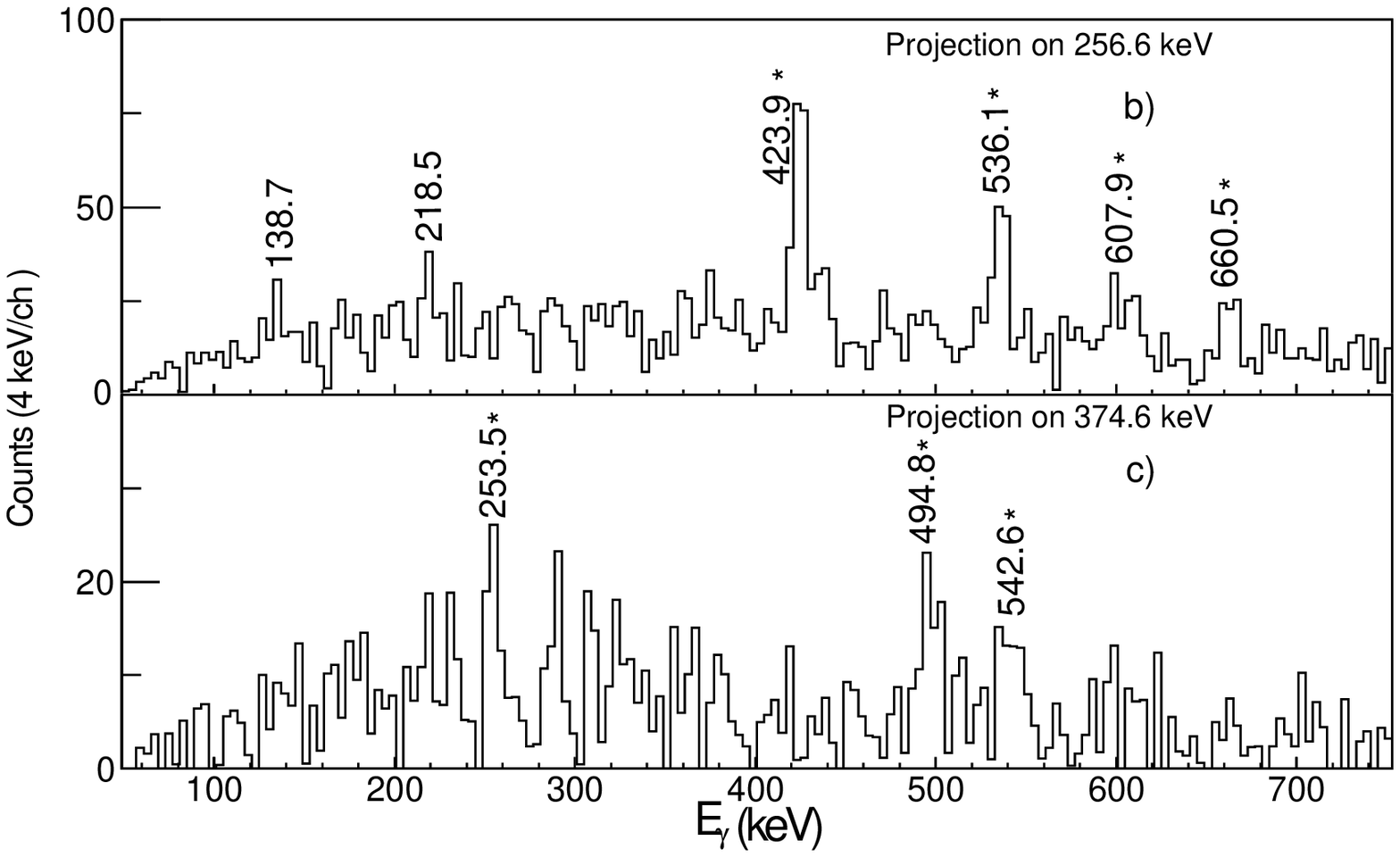}
 \caption{\label{147mass}  (a) Partial $^{147}$Pr  A- and  Z- gated
   $\gamma$-ray  spectrum obtained from  $^{238}$U +  $^{9}$Be induced
   fission data. (b)  is a spectrum gated on  the 256.6~keV transition
   and (c) is a spectrum gated on the 374.6~keV transition. The * indicates new transitions.}
\end{figure}
\begin{figure}
 \includegraphics[angle=90,width=\columnwidth]{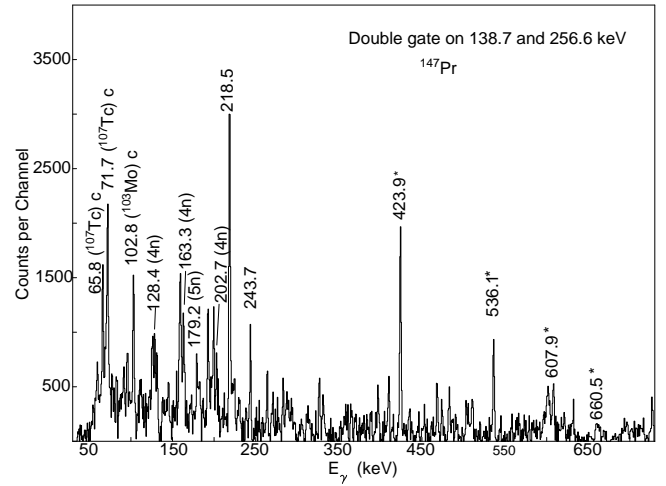}
 \caption{\label{147gate}Partial $\gamma$-ray  coincidence spectrum by
   gating  on  138.7 and  256.6~keV  transitions  in $^{147}$Pr  from
   $^{252}$Cf SF  data. In
   the  spectrum,  transitions belonging  to  Y  fission partners  are
   indicated with neutron evaporation numbers. The * indicates new transitions. }
\end{figure}

The new  level scheme for $^{147}$Pr is  shown in Fig.~\ref{147level}.
In this  case, the known  transitions were previously observed  in the
$\beta$-decay  of $^{147}$Ce  \cite{Man,Gre}. The  ground  state, 2.7,
27.9,  93.2 and  246.4~keV levels  were tentatively  proposed  to have
positive parity \cite{Man}.  The  363.3~keV level was proposed to have
a negative  parity \cite{Man}. The  (A,~Z) gated $\gamma$  spectrum is
shown in  Fig.~\ref{147mass}, where evidence  for a high spin  band is
seen.   A  new  band  is  assigned  to  $^{147}$Pr  by  observing  the
$\gamma$-rays  in  coincidence with  the  138.7,  218.5 and  243.7~keV
transitions.  A partial coincident $\gamma$-ray spectrum obtained with
$^{252}$Cf SF  data is shown  in Fig.~\ref{147gate}. In  this spectrum
with  gate  on  a   previously  known  transition  and  the  strongest
transition in Fig.~\ref{147mass}, one  can clearly see the known 218.5
and  243.7~keV  transitions, new  423.9,  536.1,  607.9 and  660.5~keV
transitions  and  the  Y  partner  transitions.   Note  the  179.2~keV
transition   labeled  in  the   spectrum  is   a  new   transition  in
$^{100}$Y. The work  including the new level scheme  of $^{100}$Y will
be published later. These  new transitions in $^{147}$Pr are confirmed
in the (A,~Z) gated  spectrum on $^{147}$Pr (Fig.~\ref{147mass}).  The
2.7  and 25.2~keV  transitions are  not observed  in the  current work
because  the  $3D$  and  $4D$  cube  data  cut  off  the  energy  from
33.3~keV. The 66 and 72~keV  transitions in this double gate come from
the coincidence of  the 138 keV transition in  $^{107}$Tc. The 103~keV
transition in this gate in Fig.~\ref{147gate} is caused by the 138 and
255~keV   transitions  in   $^{103}$Mo.   These   three  contamination
transitions are not seen in the 256.6-423.9~keV gated spectrum, so are
not in $^{147}$Pr.  The time  gated data from $^{252}$Cf SF indicate a
short lifetime  ($<$20ns) of the  385.7 keV level.  However,  since no
lifetime  of this  level was  reported  by Mantica  \emph{et al.}   in
$\beta$-decay  work~\cite{Man},  there  might  be a  very  low  energy
transition between the 138.7 and 256.6~keV transitions.  This possible
short  lifetime would  also explain  the rather  low intensity  of the
138.7~keV transition in the prompt (A,~Z) gated spectrum.  Such a loss
of intensity has already been found in the case of the neutron rich Zr
isotopes \cite{Nav13}.   The gated spectrum on  the 374~keV transition
of (A,~Z) gated  data shows evidence for a  253-374-495-542 cascade as
seen in Fig.~\ref{147mass}~(c).  This cascade is also confirmed by the
$^{252}$Cf SF data, but not  in coincidence with any other transitions
reported  in $^{147}$Ce  $\beta$-decay in  Ref.~\cite{Man,Gre}.  Thus,
level energy of this cascade is not placed in the level scheme.  Other
unlabeled transitions identified in Fig.~\ref{147mass} (155, 290, 343,
360, 435,  and 443 keV)  could not be  placed in the level  scheme for
lack of coincidence data.

Previously, internal conversion coefficient of the 138.7~keV
transition  was  not measured  in  $\beta$-decay  \cite{Man}.  In  the
current  work, $\alpha$$_{exp}$  of  the 138.7~keV  transition can  be
measured  from  the intensities  of  the  138.7,  218.5 and  243.7~keV
transitions in the 256/424 double gate with the internal conversion of
the relevant transitions included. The $\alpha$$_{exp}$ value obtained
is  0.15(3), which  is  consistent with  a  theoretical E1  transition
(0.10) but  not M1(0.51)  or E2(0.67). Thus,  band (1) is  proposed to
have a negative parity and the  band-head could be either the 362.3 or
385.1~keV level.   From the comparison with the  negative parity bands
in  $^{145}$La  \cite{Urb96,Zhu99}  and $^{149}$Pm  \cite{Jon96},  the
256.6~keV $\gamma$ ray is more likely to be a transition decaying from
15/2$^-$ to  11/2$^-$. Previously, the  ground state, 2.7,  27.9, 93.2
keV levels were tentatively  assigned to 3/2$^+$, 5/2$^+$, 7/2$^+$ and
5/2$^+$, respectively,  in $\beta$-decay work  \cite{Man} according to
the decay pattern.  The  218.5 and 269.1~keV transitions were assigned
as  M1   and  E1,  respectively,  according   to  internal  conversion
measurement \cite{Man}.   Assigning 11/2$^-$ to 385.1 keV level, the 246.4~keV level is  proposed to be
9/2$^+$. The 362.3~keV level is then 7/2$^-$ because the
11/2$^-$ level  decays to this  state. Note that 11/2$^-$  and 7/2$^-$
are   the  lowest  two   negative  levels   (<250~keV)   in  the
particle-plus-triaxial    rotor    model    (PTRM)   calculation    in
Ref.~\cite{Man}.    In contrast,   the   adopted   levels   in
Ref.~\cite{Nic} tentatively assigned the ground state, 2.7, 27.9, 93.2
and  362.3~keV  states  to  5/2$^+$,  3/2$^+$,  5/2$^+$,  7/2$^+$  and
5/2$^-$,   respectively,   based    on   the   PTRM  positive parity calculation   in
Ref.~\cite{Man}. However, this 5/2$^-$ assignment does not agree with the lowest negative parity levels as the already noted.
Thus considering both the negative and positive parity calculations, we agree with the spins and parities assigned in Ref.~\cite{Man} and not with Ref.~\cite{Nic} for the lowest states.

\subsection{$^{148}$Pr}

\begin{figure}[t]
 \includegraphics[width=0.5\columnwidth]{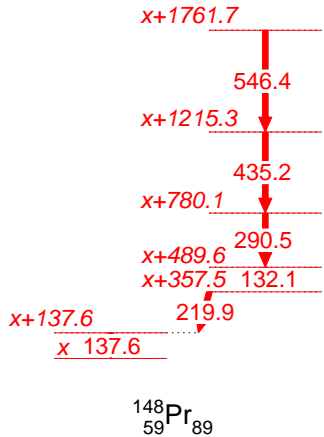}
 \caption{\label{148level}(Color   online)The  new  level   scheme  of
   $^{148}$Pr in  the present work. Relative intensities  could not be
   extracted  because  of the  very  strong  219.8  keV transition  in
   $^{149}$Pr.}
\end{figure}
\begin{figure}
 \includegraphics[width=\columnwidth]{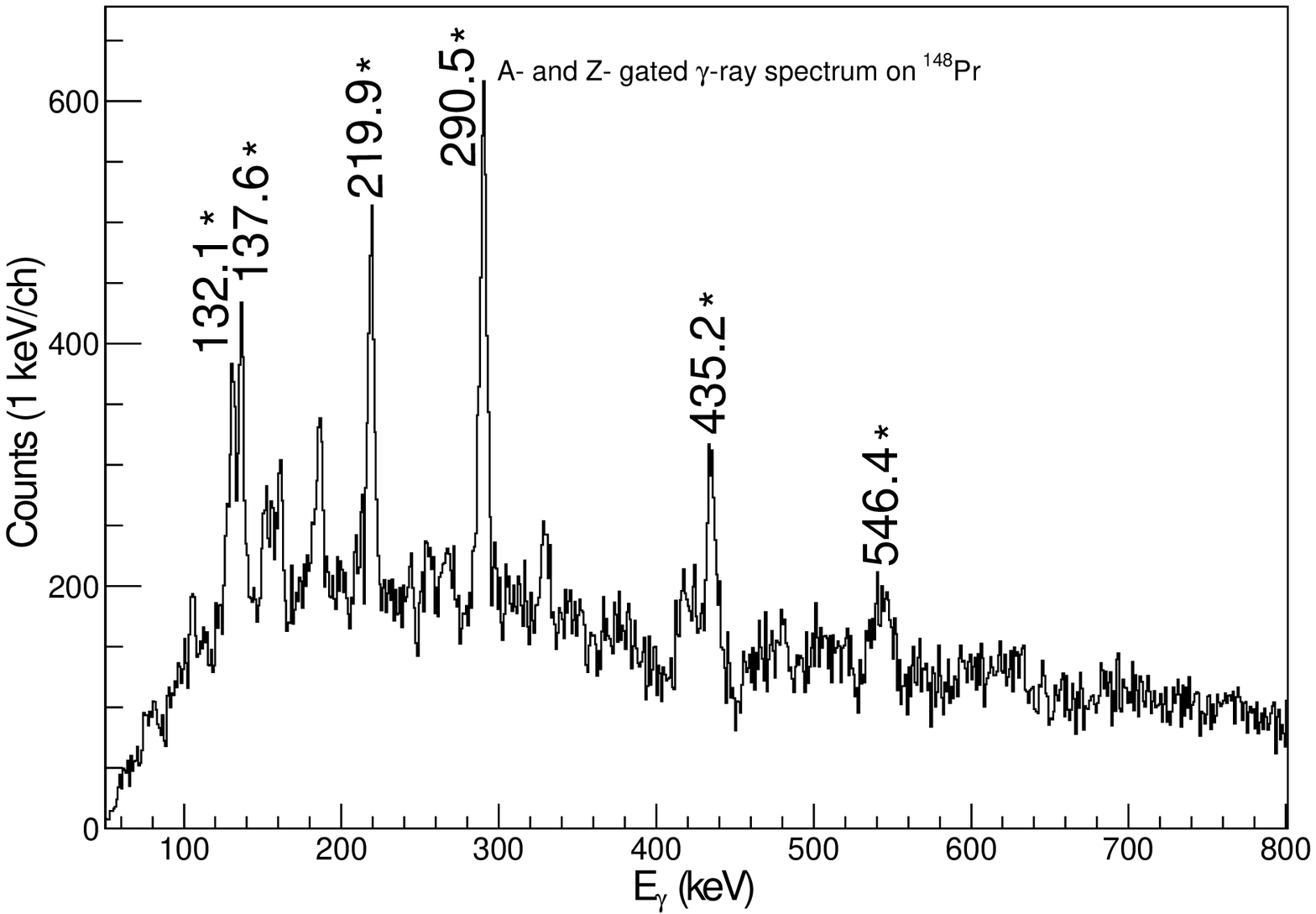}
 \caption{\label{148mass}Partial   $^{148}$Pr  mass-   and   Z-  gated
   $\gamma$-ray  spectrum obtained from  $^{238}$U +  $^{9}$Be induced
   fission data. The * indicates new transitions. }
\end{figure}
\begin{figure}
 \includegraphics[angle=90,width=\columnwidth]{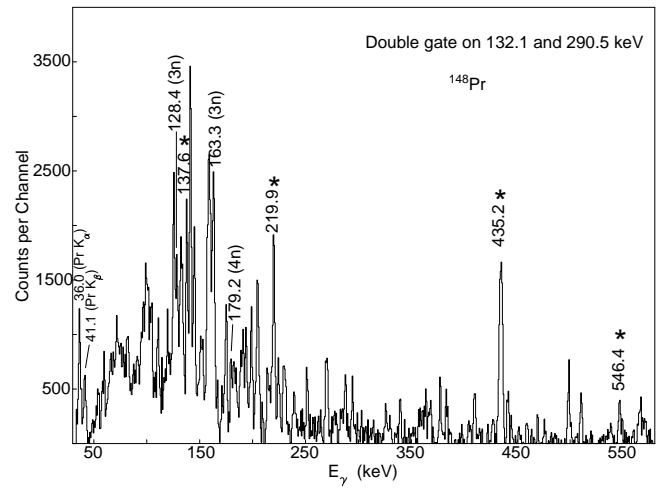}
 \caption{\label{148gate}Partial $\gamma$-ray  coincidence spectrum by
   gating  on  132.1 and  290.5~keV  transitions  in $^{148}$Pr  from
   $^{252}$Cf SF  data. In
   the  spectrum,  transitions belonging  to  Y  fission partners  are
   indicated with the corresponding number of evaporated neutrons. The * indicates new transitions.}
\end{figure}

The new  level scheme for $^{148}$Pr is  shown in Fig.~\ref{148level}.
In  this   case,  all  transitions  are  newly   identified.  The  new
transitions  are  seen in  the  mass-Z  gated  spectrum of  $^{148}$Pr
(Fig.~\ref{148mass}).  None  of transitions from  the $\beta$-decay of
$^{148}$Ce  \cite{Gre,Ara,Koj} are  observed in  this work.  A partial
coincident $\gamma$-ray spectrum from  the $^{252}$Cf data is shown in
Fig.~\ref{148gate}.   In  this  spectrum   with  gates  on  132.1  and
290.5~keV transitions, one can clearly see the 137.6, 219.9, 435.2 and
546.4~keV new transitions and the  Y partner transitions. The order of
these  new transitions  are placed  based on  the intensities  and the
similarity to $^{150}$Pr.  In $^{252}$Cf SF data, the intensity ratios
of  137.6/132.1 and  219.9/132.1  decrease about  80$\%$  as the  time
coincidence window decreases  from 500ns to 8ns. In  the contrast, the
ratio of the 132.1,  290.5, 435.2 and 546.4~keV transition intensities
remain almost the same.  Thus, the 219.9 and 137.6~keV transitions are
placed at  the bottom and a  lifetime of the order of a hundred ns is proposed for  the level which
the  132.1~keV  transition feeds.  This  is  consistent  with the  non
observation of the 219.9~keV  transition in coincidence with 290.5~keV
in the  prompt $\gamma$ ray  spectrum obtained using the  (A,~Z) gate where the prompt gamma rays are sensitive only to states with lifetimes  shorter than $\sim$2 ns..
In addition spectra obtained in A/Z coincidence with 219.9~keV suggest the
existence of another band consisting of 219.9 and 329~keV transitions. But we cannot confirm this cascade in SF data.
Triple gates (137.6-132.1-290.5 and 219.5-132.1-290.5 keV) show no evidence for the strong 105, 121, 195 or 289 keV transitions seen in $\beta$-decay.
Thus we conclude none of the low spins (1,2) states seen in $\beta$-decay are feed by the cascade seen in Fig.~\ref{148mass}. This cascade could feed the
2 mins 4$^-$ isomer at 77 keV \cite{Koj}.

\subsection{$^{149}$Pr}

\begin{figure}
 \includegraphics[width=\columnwidth]{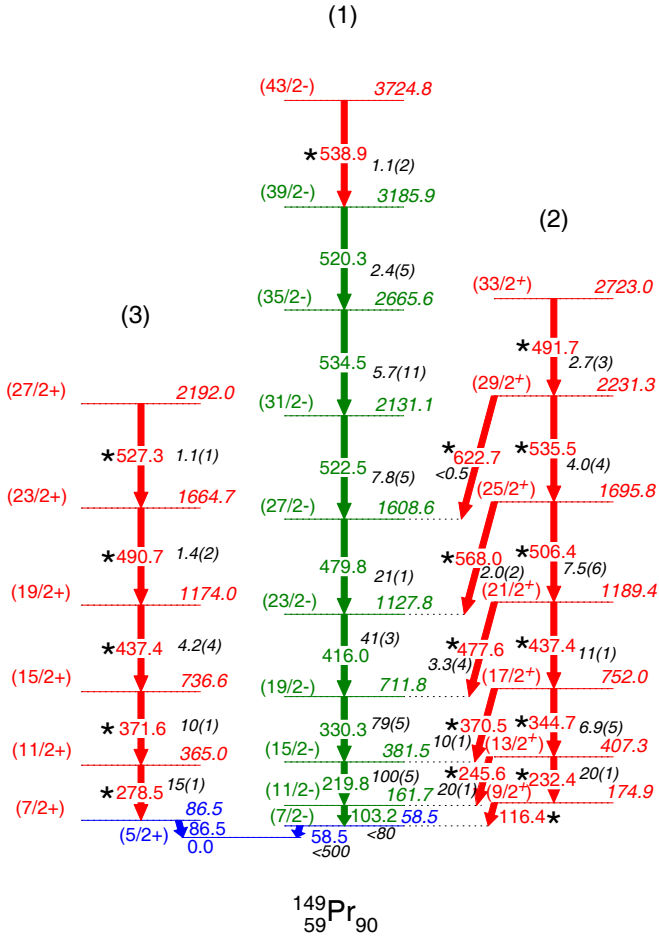}
 \caption{\label{149level}(Color   online)The  new  level   scheme  of
   $^{149}$Pr in  the present work. The 86.5 and 58.5 keV transitions  and levels previously
   reported in $\beta$ decay work  are labeled in blue. Those reported
   by  Hwang {\it  et  al.}  \cite{Hwa00} are  labeled  in green, the 103.2 to 520.3 keV transitions in band (1).  New
   transitions and levels are labeled in red and with an asterisk. The intensities in black to the right of the $\gamma$ ray energies are relative to 100 for the 219.8 keV transition.}
\end{figure}
\begin{figure}
 \includegraphics[width=\columnwidth]{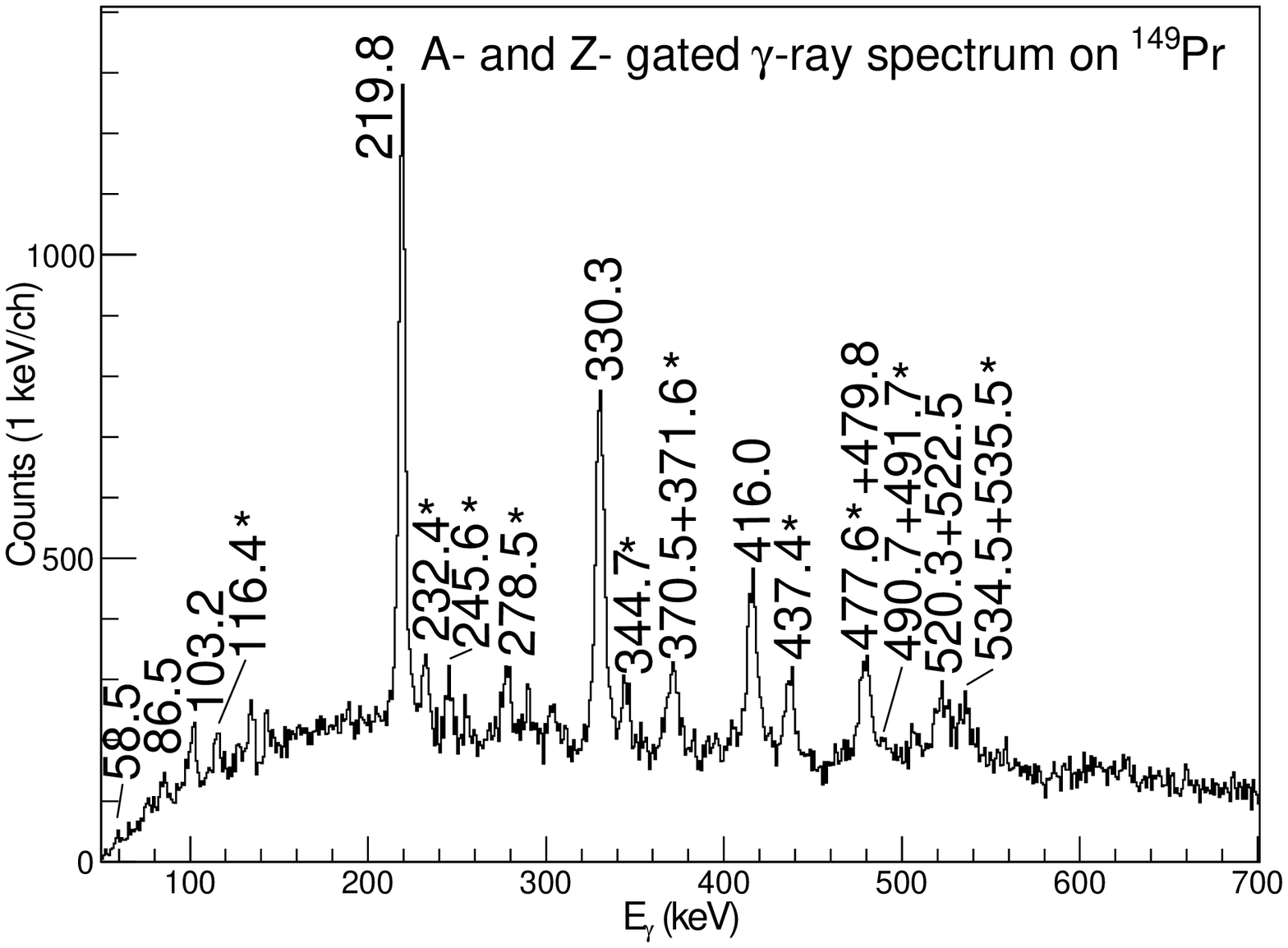}
 \caption{\label{149mass}Partial   $^{149}$Pr    A-   and   Z-   gated
   $\gamma$-ray  spectrum obtained from  $^{238}$U +  $^{9}$Be induced
   fission data. The * indicates new transitions. }
\end{figure}
\begin{figure*}
 \includegraphics[width=\columnwidth,angle=90]{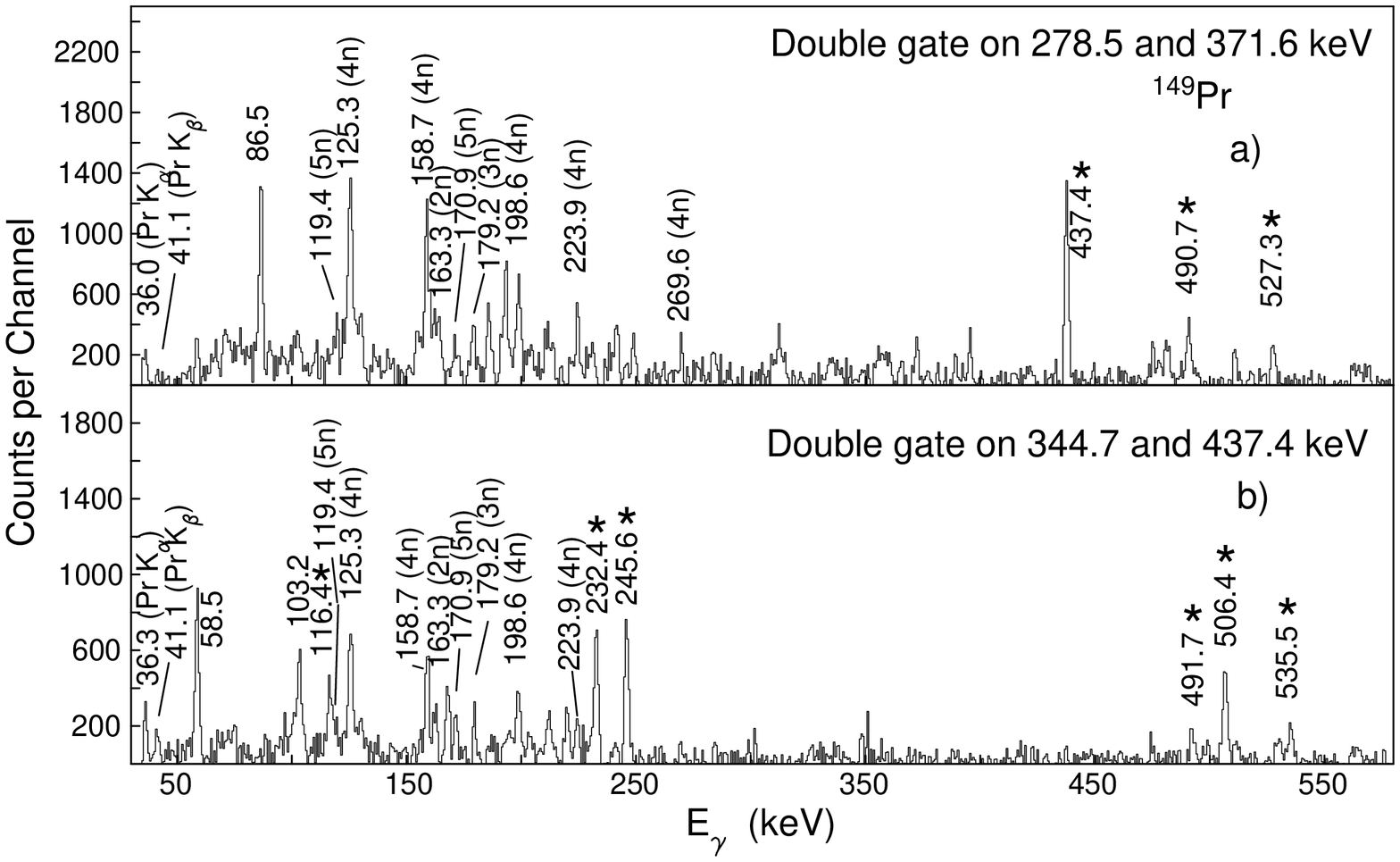}
 \caption{\label{149gate}Partial $\gamma$-ray  coincidence spectra (a)
   by gating on 278.5 and 371.6~keV transitions, and (b) by gating on
   344.7 and  437.4~keV transitions  in $^{149}$Pr from  $^{252}$Cf SF
   data. In
   the  spectrum,  transitions belonging  to  Y  fission partners  are
   indicated with  neutron evaporation numbers,  specifically, 5n, 4n,
   3n, 2n correspond to $^{98,99,100,101}$Y respectively. The * indicates new transitions. }
\end{figure*}

\begin{figure}
 \includegraphics[width=\columnwidth]{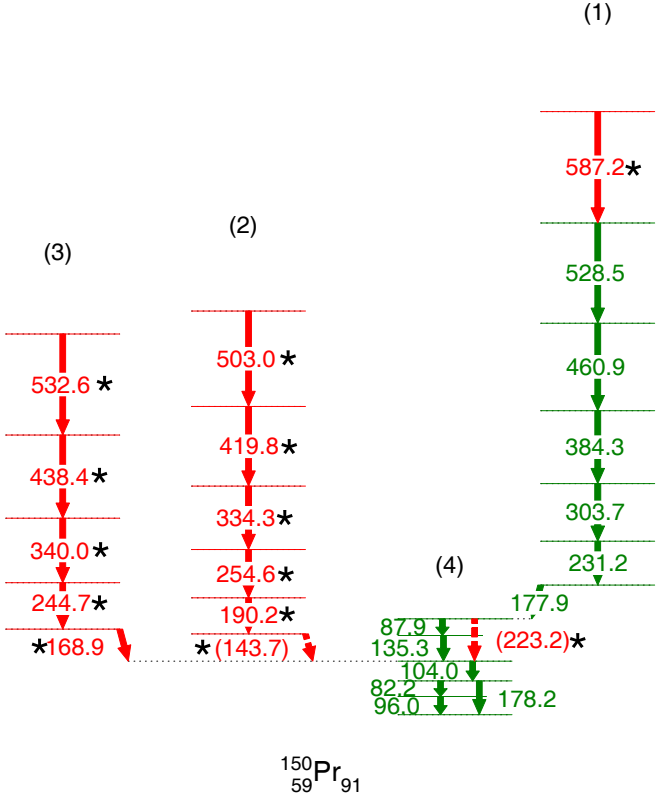}
 \caption{\label{150level}The  new level scheme  of $^{150}$Pr  in the
   present  work.  Transitions   and  levels  previously  reported  by
   Hwang\cite{Hwa00,Hwa10} are labeled in  green. New ones are labeled
   in red and with an asterisk.}
\end{figure}
\begin{figure}
 \includegraphics[width=\columnwidth]{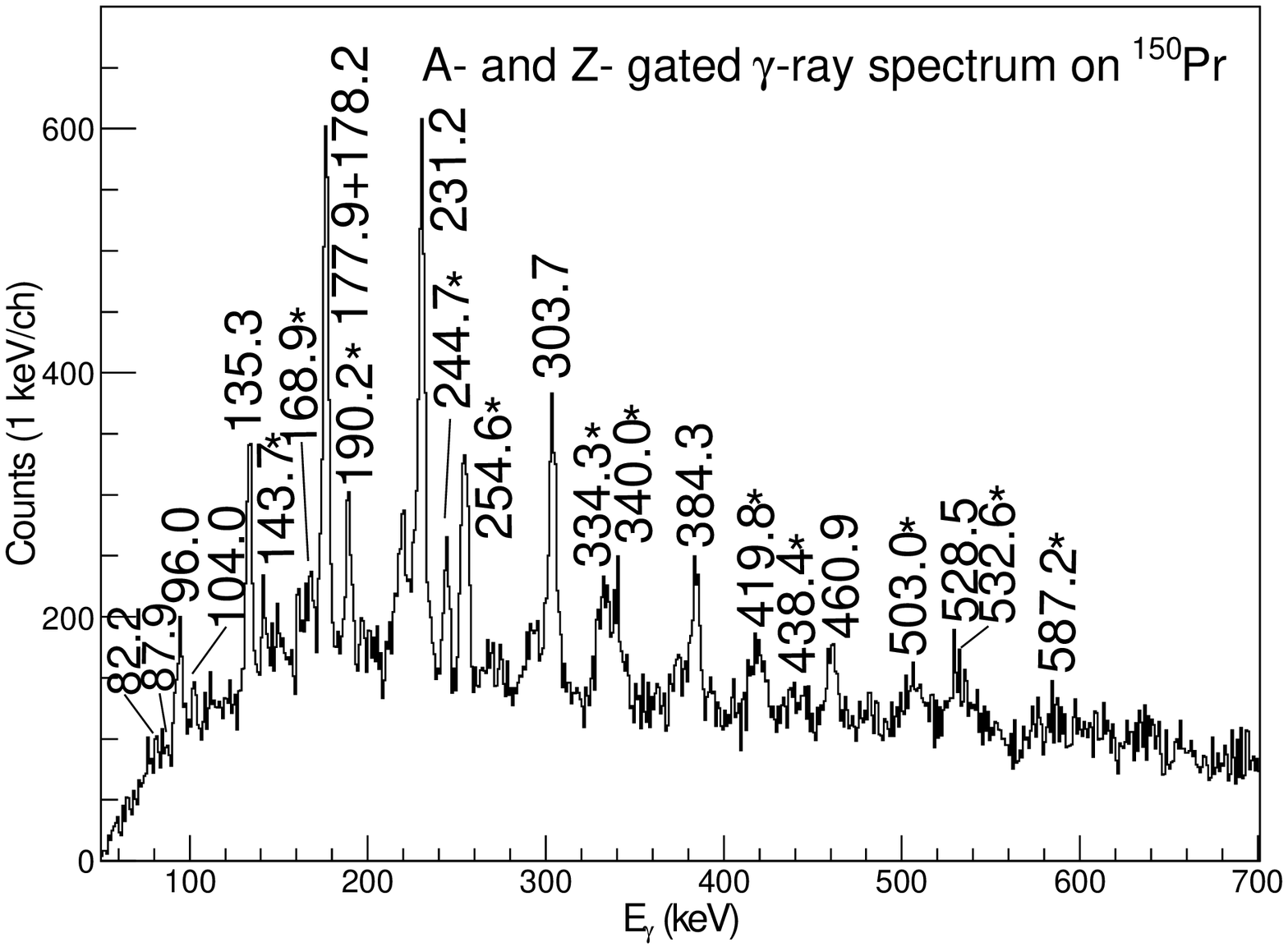}
 \caption{\label{150mass}Partial   $^{150}$Pr  mass-   and   Z-  gated
   $\gamma$-ray  spectra obtained  from $^{238}$U  +  $^{9}$Be induced
   fission data. The * indicates new transitions. }
\end{figure}

\begin{figure*}
 \includegraphics[width=\columnwidth,angle=90]{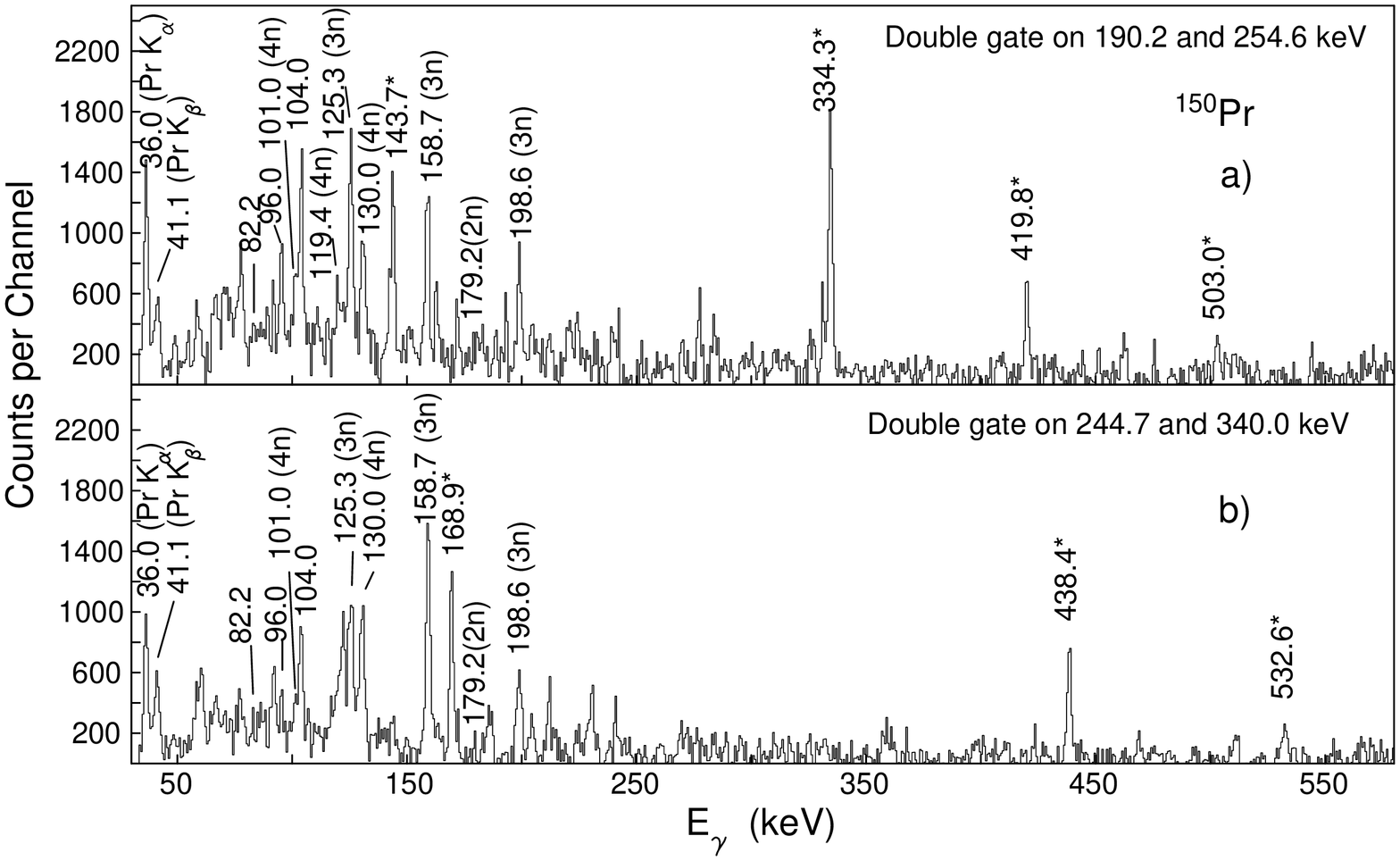}
 \caption{\label{150gate} Partial $\gamma$-ray coincidence spectra (a)
   by gating on 190.2 and 254.6~keV transitions, and (b) by gating on
   244.7  and  340.0  transitions  in $^{150}$Pr  from  $^{252}$Cf  SF
   data. In
   the  spectrum,  transitions  belonging  to Y  fission  partner  are
   indicated with  neutron evaporation numbers,  specifically, 4n, 3n,
   2n correspond to $^{98,99,100}$Y, respectively. The * indicates new transitions. }
\end{figure*}

The    new    level    scheme    for   $^{149}$Pr    is    shown    in
Fig.~\ref{149level}. In  this case, the 58.5  and 86.5~keV transitions
were   previously  observed   in  the   $\beta$-decay   of  $^{149}$Ce
\cite{Pfe,Keys}.  The  band (1) in  Fig.~\ref{149level} was previously
reported  in  Ref.~\cite{Hwa00}.   In  Fig.~\ref{149mass},  previously
reported band (1) transitions and  new transitions in band (2) and (3)
are seen in the (A,~Z) gated spectra.  These spectra were important in
guiding SF $\gamma$ ray coincidence  spectra analysis to identify the new
band (3).   These two  new bands  (2) and (3)  are established  in the
present work by observing the coincidence between the $\gamma$-rays in
these bands and the 86.5 or 58.5~keV transition in $^{252}$Cf SF data.
A   partial    coincident   $\gamma$-ray   spectrum    is   shown   in
Fig.~\ref{149gate}(a).   In  this spectrum  with  a  gate  on two  new
transitions in band (3), one  can clearly see the 86.5~keV transition,
three other new transitions  and Y partner transitions. Another partial  coincident $\gamma$-ray spectrum
is shown in  Fig.~\ref{149gate}(b). In this spectrum with  gate on two
new transitions  in band (2) one  can clearly see  the 58.5, 103.2~keV
transitions, six  other new transitions and Y  partner transitions. By
gating   on  these   new  transitions   and  analyzing   the  relative
$\gamma$-transition intensities, these two  new bands are proposed for
$^{149}$Pr. The $^{252}$Cf data also  shows some weak evidence for the
coincidences  between  the  103.2   and  232.4,  219.8  and  344.7~keV
transitions,  respectively. The  possible  low energy  13.2 (174.9  to
161.7) and 25.8 (407.3 to  381.5)~keV transitions are not indicated in
the level scheme in Fig.~\ref{149level}.

Spins and parities of levels  in band (1) were tentatively assigned in
Ref.~\cite{Rza} by an  internal conversion coefficient measurement and
theoretical   calculations.  Those   results  are   adopted   in  this
paper.  Based   on  the  regular  energy   spacings  and  $\gamma$-ray
intensities, the new level at 3724.8~keV is assigned as 43/2$^-$.

The total  internal conversion coefficient of the  86.5~keV transition
in $^{149}$Pr was  measured from the intensity ratio  between 86.5 and
278.5~keV  transition in  the coincident spectrum  gated on  371.6 and
437.4~keV  transitions above  them.  The  value  was obtained  to  be
1.63(22), and  is in agreement  with the theoretical  calculated value
\cite{bri} of  1.96 for a  M1 transition but  not with 3.58 for  an E2
transition.

Quasiparticle-rotor model (QPRM) calculations \cite{Rza,Gab} suggest a
$\pi$5/2[413] configuration for the $^{149}$Pr ground state level. The
lowest three exited  level energies at 86.5, 365.0,  736.6~keV of band
(3)  in  $^{149}$Pr  are  also  reasonably consistent  with  the  QRPM
calculations of 7/2$^+$,  11/2$^+$ and 15/2$^+$ for the  states at 87,
342 and 708~keV, respectively in Ref.~\cite{Rza}. Therefore, the spins
and parities of exited states  in band (2) are tentatively assigned as
7/2$^+$, 11/2$^+$,  15/2$^+$, 19/2$^+$,  23/2$^+$ and 27/2$^+$  in the
present work.

Spins and parities of the  levels in band (2) are tentatively assigned
in  Fig.~\ref{149level}  based  on  the structure  similarity  to  the
$h_{11/2}$   signature   s=+i   octupole   bands  (1)   and   (2)   in
Ref.~\cite{Zhu99}.  When octupole deformation or octupole correlations
are strong, one expects a  symplectic quantum number s=i, two bands of
opposite parity  with the spin show in  Fig.~\ref{149level} and strong
E1 transitions between the two  bands as found \cite{Naza}.  Note that
QPRM  calculation  in  Ref.~\cite{Rza}  indicates that  this  band  is
unlikely  to be  another signature  of band  (1). Further  analysis is
included in the discussion part.

\subsection{$^{150}$Pr}

The new  level scheme for $^{150}$Pr is  shown in Fig.~\ref{150level}.
Bands  (1) and  (4) were  previously  assigned to  $^{150}$Pr in  Ref.
\cite{Hwa00,Hwa10} from  the SF of  $^{252}$Cf. In this  earlier work,
the  relative  yield  ratios  of  partner  Y  isotopes  were  measured
\cite{Hwa10}.  The transitions previously assigned to $^{150}$Pr along
with  several  new  ones  are  shown  in  the  (A,~Z)  gated  spectrum
(Fig.~\ref{150mass}).    Fig.~\ref{150gate}(a)  shows   a  coincidence
spectrum    double-gated   on    the   new    190.2    and   254.6~keV
transitions. Fig.~\ref{150gate}(b) shows a coincidence spectrum double
gated on the new  244.7 and 340.0~keV transitions.The previously known
$\gamma$-transitions of  96, 82.2 and 104.0~keV of  $^{150}$Pr and the
$\gamma$-transitions in  the partner Y  isotopes can be seen  in these
spectra. Therefore,  these new transitions are  assigned to $^{150}$Pr
in the  present work.  The newly observed  334.3, 419.8  and 503.0~keV
transitions  are   coincident  with   the  new  190.2   and  254.6~keV
transitions  in   Fig.~\ref{150gate}(a).  Also,  the   new  438.4  and
532.6~keV transitions are coincident  with the new 340.7 and 244.7~keV
transitions  in  Fig.~\ref{150gate}(b).   By using  these  coincidence
relationships in multiple gates, bands (2) and (3) were found as shown
in  Fig.~\ref{150level}. The 143.7~keV  transition is  dashed because
the intensity is weaker than that of the 190.2~keV transition.

\subsection{$^{151}$Pr}

\begin{figure}
 \includegraphics[width=\columnwidth]{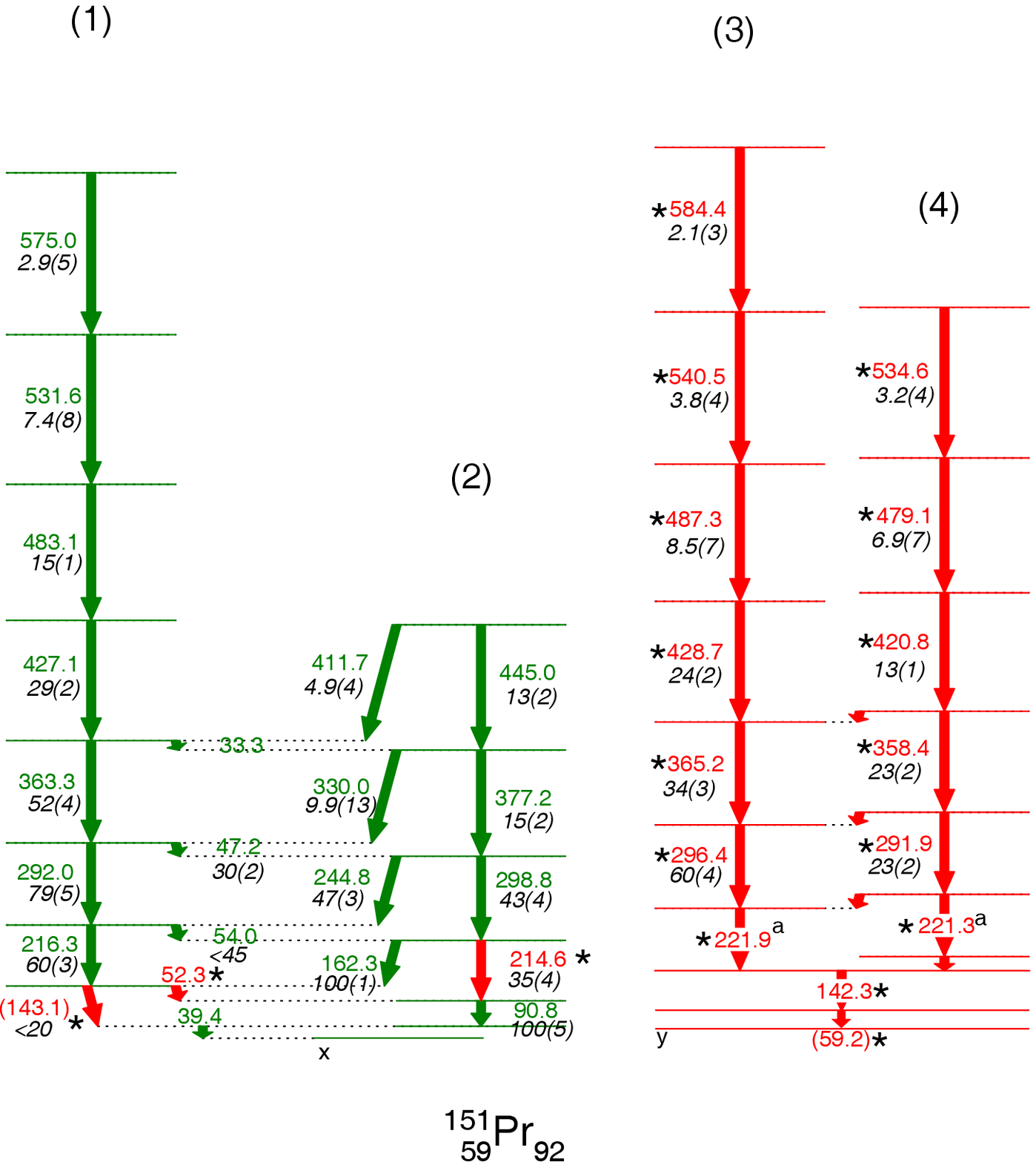}
 \caption{\label{151level} The  new level scheme of  $^{151}$Pr in the
   present work.  Transitions and  levels previously reported by Hwang
   {\it  et al.}\cite{Hwa10}  are labeled  in green.   New  levels and
   transitions  are   labeled  in  red  and  with   an  asterisk.  The
   221.9+221.3~keV  transitions, superscript  a,  have the  normalized
   intensity of 100. Band (3) and  (4) are shown in red with other new
   transitions  since   both  band  now  feed   a  'single'  142.3~keV
   transition.  Band  (3) was assigned  to $^{151}$Pr and band  (4) to
   $^{153}$Pr in  Ref.~\cite{Hwa10}. The $\gamma$-ray  intensities are
   relative to 100 for the 162.3 keV transition.}
\end{figure}
\begin{figure}
 \includegraphics[width=\columnwidth]{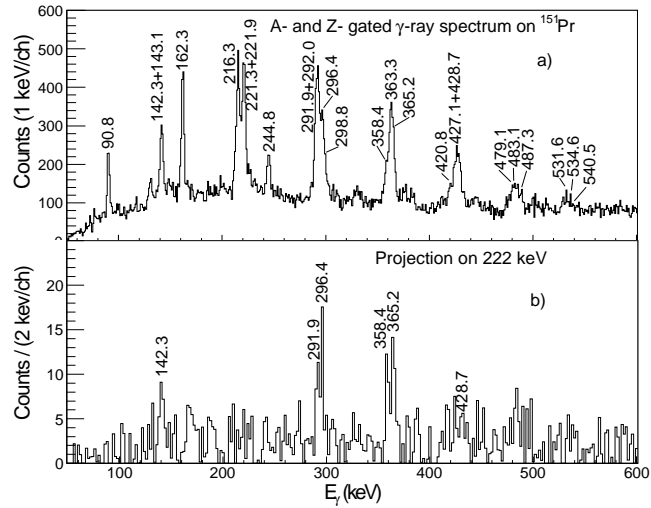}
 \caption{\label{151mass}Partial   $^{151}$Pr  mass-   and   Z-  gated
   $\gamma$-ray  spectra obtained  from $^{238}$U  +  $^{9}$Be induced
   fission data.   Part (a) is 'single' $\gamma$-ray  spectrum. Part (b)
   is a gated spectrum on the 222~keV transition.}
\end{figure}
\begin{figure*}
 \includegraphics[width=\columnwidth,angle=90]{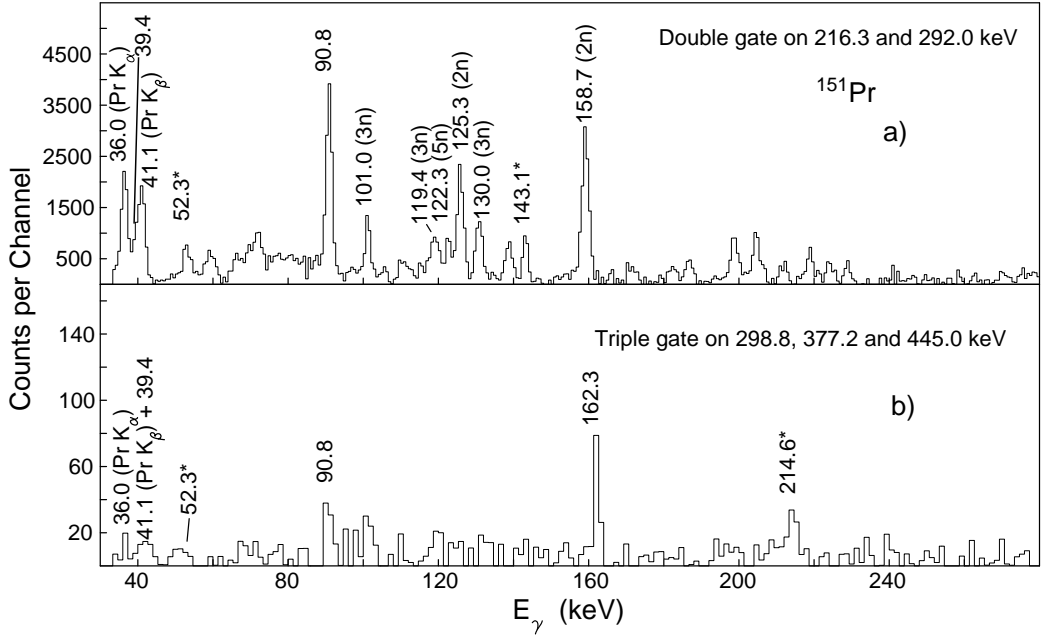}
 \caption{\label{151gate}Partial $\gamma$-ray  coincidence spectra (a)
   by gating on 216.3 and 292.0~keV transitions, and (b) by gating on
   298.8,  377.2   and  445.0~keV  transitions   in  $^{151}$Pr  from
   $^{252}$Cf SF  data. In the  spectrum,  transitions belonging  to Y  fission
   partner   are   indicated   with   neutron   evaporation   numbers,
   specifically, 5n, 3n, 2n correspond to $^{96,98,99}$Y respectively. The * indicates new transitions. Note the keV/channel in part a) differs from part b).}
\end{figure*}

\begin{figure*}
 \includegraphics[width=1.5\columnwidth]{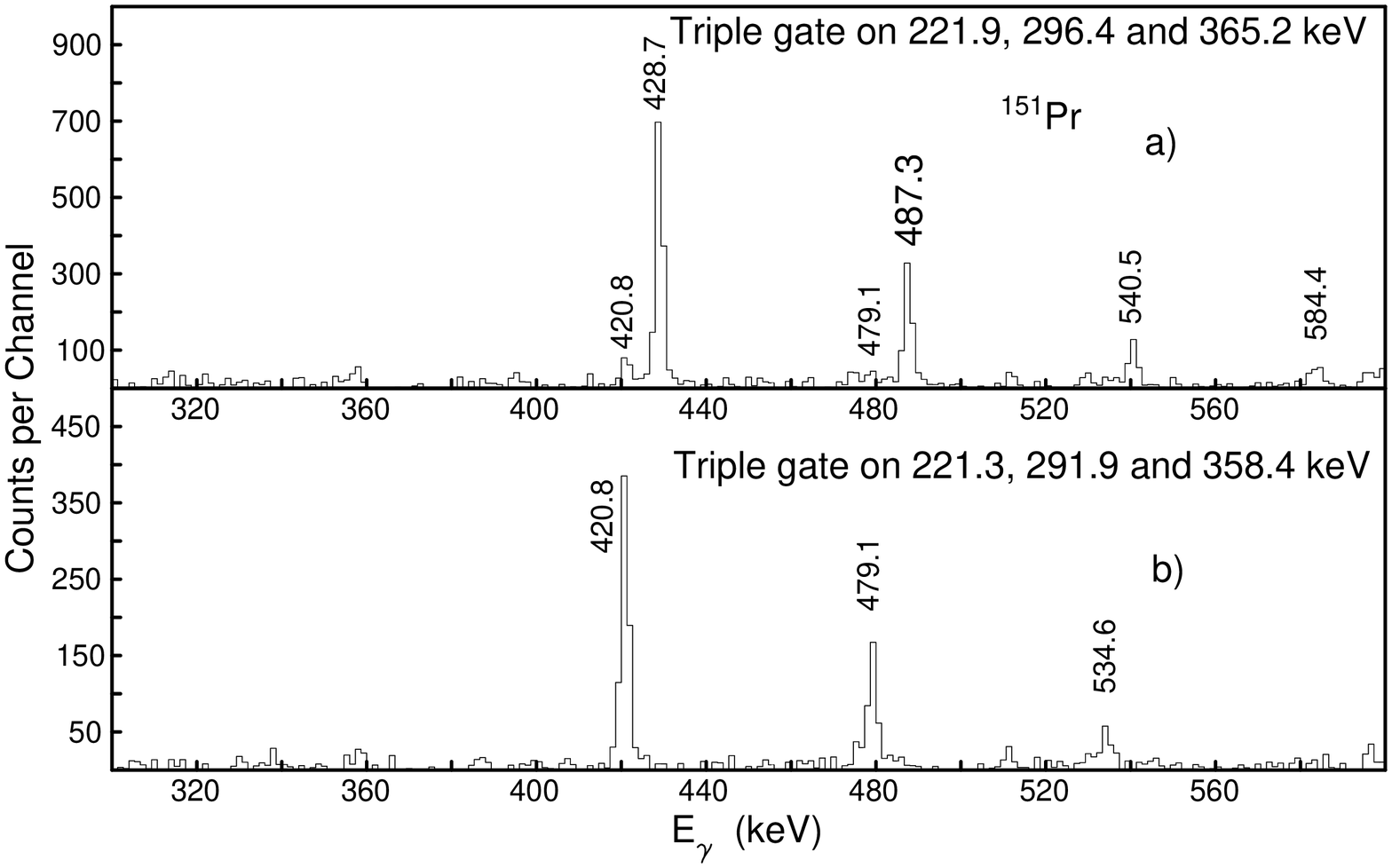}
 \caption{\label{151gate2}Partial $\gamma$-ray coincidence spectra (a)
   by gating on  221.9, 296.4 and 365.2~keV transitions,  and (b) by gating
   on 221.3, 291.9 and 358.4~keV transitions in $^{151}$Pr from $^{252}$Cf SF
   data.}
\end{figure*}
The    new    level    scheme    for   $^{151}$Pr    is    shown    in
Fig.~\ref{151level}.  The $^{151}$Pr  A and  Z gated  spectra obtained
from   $^{238}$U   +   $^{9}$Be   induced   fission   are   shown   in
Figs.~\ref{151mass}(a)    and   (b).    Bands   (1)    and    (2)   in
Fig.~\ref{151level}   were  assigned   previously  to   $^{151}$Pr  in
Ref.~\cite{Hwa10}.  In this earlier work, the relative yield ratios of
partner  Y isotopes  were measured.  However, bands  (1) and  (2) were
assigned to  $^{152}$Pr, by  Malkiewicz \emph{et al.}  \cite{Mal}. The
previously reported \cite{Hwa10} $\gamma$-transitions in bands (1) and
(2)   are  confirmed   from  the   mass  and   Z  gated   spectrum  in
Fig.~\ref{151mass}(a).  In the present  work, the  204.2 and  41.9~keV
transitions in Ref.~\cite{Hwa10} are replaced  with the new 214.6 and  52.3~keV transitions
in  bands (1)  and (2)  of Fig.~\ref{151level}.  Also, a  possible new
143.1~keV transition  is added.  Previously, bands  (3) and  (4) were
assigned to $^{152}$Pr in Ref. \cite{Liu}.

Then band  (3) was assigned to  $^{151}$Pr and band  (4) to $^{153}$Pr
from the  SF work of  $^{248}$Cm and $^{252}$Cf \cite{Mal}.   Now band
(3) and band  (4) in Fig.~\ref{151level} are assigned  to $^{151}$Pr in
the present  work because the 142.3, 221.3+221.9,  296.4, 291.9, 365.2
and 358.4~keV transitions are seen  in the $^{151}$Pr mass and Z gated
spectra in Figs.~\ref{151mass}(a)  and \ref{151mass}(b). The 221.3 and
221.9~keV  transitions  were   reported  earlier  as  one  221.9~keV
transition   \cite{Liu}.   Later,   the   142.3~keV   transition   in
Ref.~\cite{Liu} was separated into  142.1 and 141.6~keV transitions in
Ref.~\cite{Mal}.  The  221.9~keV  transition in  Ref.~\cite{Liu}  was
separated into 221.8 and  221.0~keV transitions in Ref.~\cite{Mal}. In
this paper, a  shift of about 0.6~keV of the 221.9~keV  peak has been
confirmed  by comparing  the  gates  between bands  (3)  and (4),  for
example, double  gates on 296.4 and  356.2~keV in band  (3), 291.9 and
358.4~keV  in band (4), triple  gates on 296.4,356.2 and  428.7~keV in
band  (3), 291.9,  358.4 and  420.8~keV  in band  (4). But  no visible
energy difference of the 142.3~keV transition is seen when comparisons
are  set between any  of the  gates in  bands (3)  and (4).  Thus, two
different 221.9  and 221.3~keV  transitions are proposed but  there is
only one 142.3~keV transition in the level scheme. In the present work
all  of  bands (1),  (2),  (3)  and (4)  are  definitely  assigned  to
$^{151}$Pr, as  shown Fig.~\ref{151level}. Further  analysis about the
mass assignment will be presented in the $^{152,153}$Pr and discussion
sections.

Fig.~\ref{151gate}(a)  is a  coincidence spectrum  from  $^{252}$Cf SF
data by gating on 216.3 and 292.0~keV transitions showing the new 52.3
and 143.1~keV  transitions. The peak around 40~keV is  an overlap of a
39.4~keV  $\gamma$ transition and  41.0~keV Pr X-ray.   Therefore, the
energy of the 39.4~keV transition has a relative large uncertainty and
might range from 39 to 42~keV.  If the 39.4~keV transition is the same
as the 38.9~keV one reported in $\beta$-decay  work \cite{Koj}, bands
(1) and (2) would decay to the ground state. However, since the energy
of this transition has a large  uncertainty, an x~keV level is used in
Fig.~\ref{151level}. The 143.1~keV  transition is much weaker than the
90.8~keV    one   and    is   bracketed    in   the    level   scheme.
Fig.~\ref{151gate}(b)  is a  coincidence spectrum  from  $^{252}$Cf SF
data  triple gated  on 298.8,  377.2 and  445.0~keV  transitions.  The
204.2~keV transition  is not seen compared to 214.6~keV in this gate,
which suggest some  contamination at 204.2~keV in  the spectrum of the
previous work  (Ref.~\cite{Hwa10}). Thus, the 204.2~keV transition is
replaced by the 214.6 one in the present work.

Fig.~\ref{151gate2}(a)  is a coincidence  spectrum from  $^{252}$Cf SF
data by gating  on 221.9, 296.4 and 365.2~keV  transitions. The 428.7,
487.3, 540.5  and 584.4~keV transitions  in band (3)  can be  seen.
 Fig.~\ref{151gate2}(b) is a coincidence spectrum from $^{252}$Cf
Sf  data by  gating on  221.3, 291.9 and 358.4~keV  transitions.  The
420.8, 479.1  and 534.6~keV transitions  in band (4)  can be  seen. The $^{252}$Cf SF
data also show coincidence of the  296.4 and 358.4~keV transitions, as well
as  the 365.2  and 420.8~keV transitions,  but  interband transitions
linking bands (3) and (4) are  not observed in the current work. These
coincidences indicate the existence of a very low energy transition to
lift up band (4) in $^{151}$Pr.

The  analysis  result for  $\gamma$  transition  intensities are  also
labeled in the level scheme.  The intensities have been separated into
two parts. The intensities of transitions in band (1) and band (2) are
normalized to that  of the 162.3~keV transition and  those in band (3)
and band  (4) are normalized to  the summation of the  221.9 and 221.3
keV transition intensities.

Previously, internal conversion  measurements established the 47.2 and
54.0~keV  transitions as E1 and  90.8~keV one as  M1 \cite{Hwa10}. The
value of the 90.8~keV transition  was obtained from the 90.8 and 292.0
keV transition intensities in  the coincidence spectrum gated on 216.3
and 363.3~keV in Ref.~\cite{Hwa10}. This measurement  did not include
the contribution of  the 47.2~keV transition. If  this contribution is
taken into account,  the internal conversion value will  increase by a
factor of $\sim$20\%, which will  make it closer to the theoretical M1
value.   Those    corrections   can    be   seen   in    the   erratum
Ref.~\cite{Hwaer}. The  branching of the 143.1~keV  transition is very
small  (see  Fig.~\ref{151gate}(a)) and  does  not  change the  result
much. The $\alpha$$_{exp}$ value of the 54.0~keV was obtained from the
54.0,  162.0 and  204.0~keV  transition intensities  in the  292.0 and
363.3~keV  double gate in Ref.~\cite{Hwa10}. This  measurement is also
questionable because the 204.0~keV  transition is not confirmed in the
current work.  The peak  around 54.0~keV in the  292.0 and  363.3~keV
double gate is  a overlap of the 52.3 and  54.0~keV transitions. Thus,
the intensity  as well as the  $\alpha$$_{exp}$ value of  the 54.0~keV
transition  cannot be  accurately obtained  in the  present  work. The
$\alpha$$_{exp}$ value of  the 52.3~keV can be  obtained from the 52.3
and 90.8~keV transition intensities  in the 216.3 and 292.0~keV double
gate. The value is 3.5(9), which lies between the theoretical value of
E1(1.4) and M1(8.3). However, the  216/292 gate might be interfered by
the 214.6~keV transition  so that the  error of  the $\alpha$$_{exp}$
value of the 52.3~keV would be much higher.

\begin{figure}
 \includegraphics[width=0.8\columnwidth]{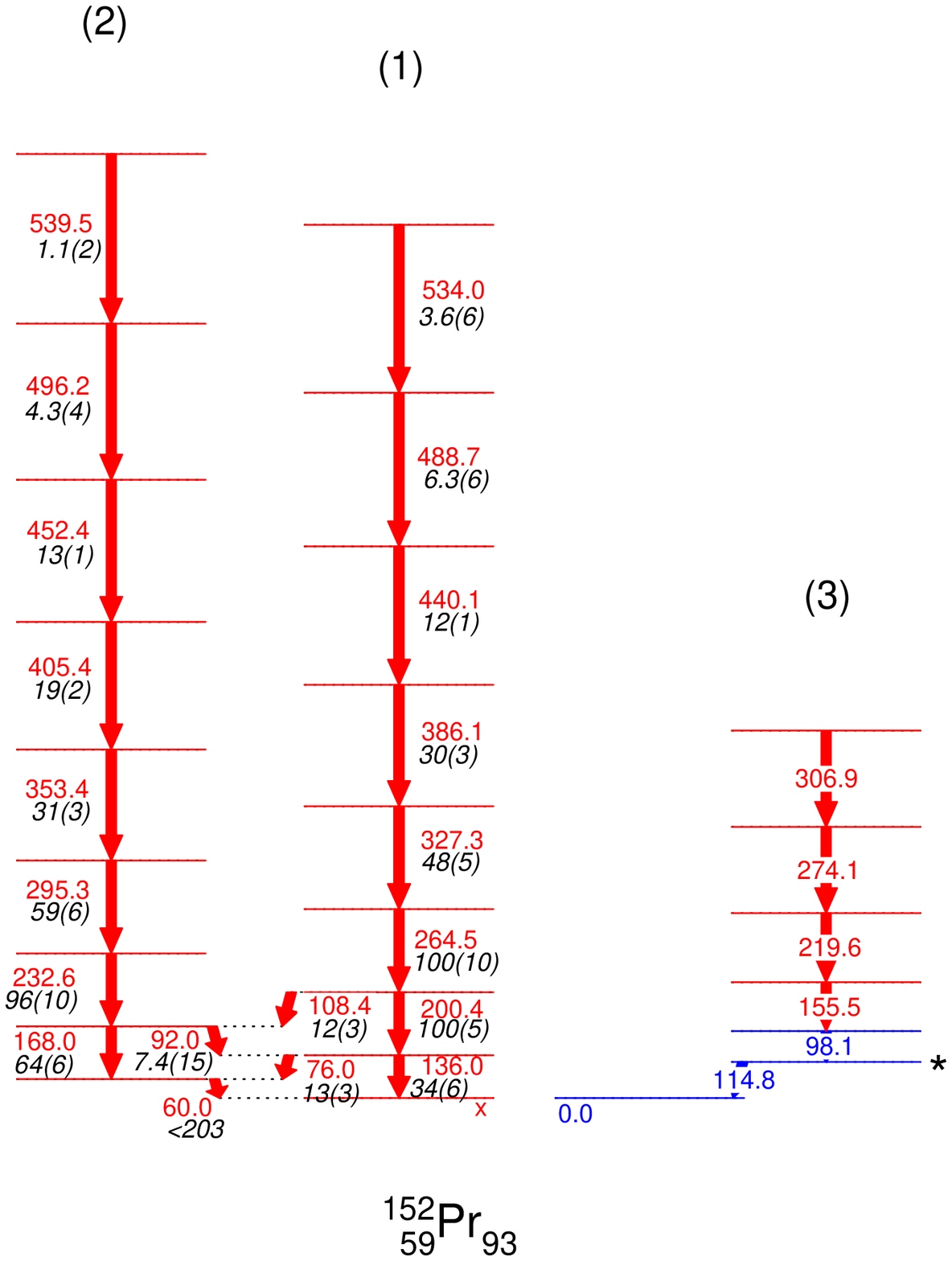}
 \caption{\label{152level}The  level  scheme   of  $^{152}$Pr  in  the
   present work. The 114.8 and 98.1 keV transitions and levels previously reported in $\beta$
   decay work are labeled in blue. All others are new ones labeled in red. * This
   level is  an isomer with a  lifetime of 4.1$\mu$s  according to the
   measurement in Ref.~\cite{Yam} The $\gamma$ ray intensities are relative to 100 for the 200.3 keV transition.}
\end{figure}
\begin{figure}
\includegraphics[width=\columnwidth]{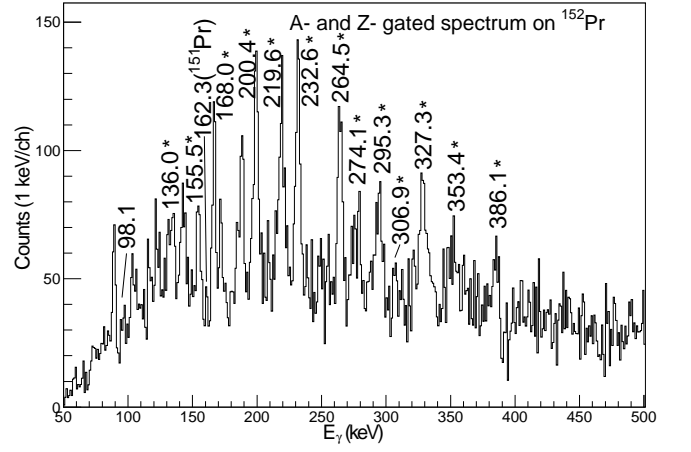}
 \caption{\label{152mass}Partial   $^{152}$Pr  mass-   and   Z-  gated
   $\gamma$-ray  spectra obtained  from $^{238}$U  +  $^{9}$Be induced
   fission  data.   The position  of  the  162.3  keV transition  from
   $^{151}$Pr is also indicated to illustrate its non-observation (see
   text  for details).   Note the  absence of  the  114.8~keV isomeric
   transition. The * indicates new transitions. }
\end{figure}
\subsection{$^{152}$Pr}

\begin{figure*}
  \includegraphics[width=\columnwidth,angle=90]{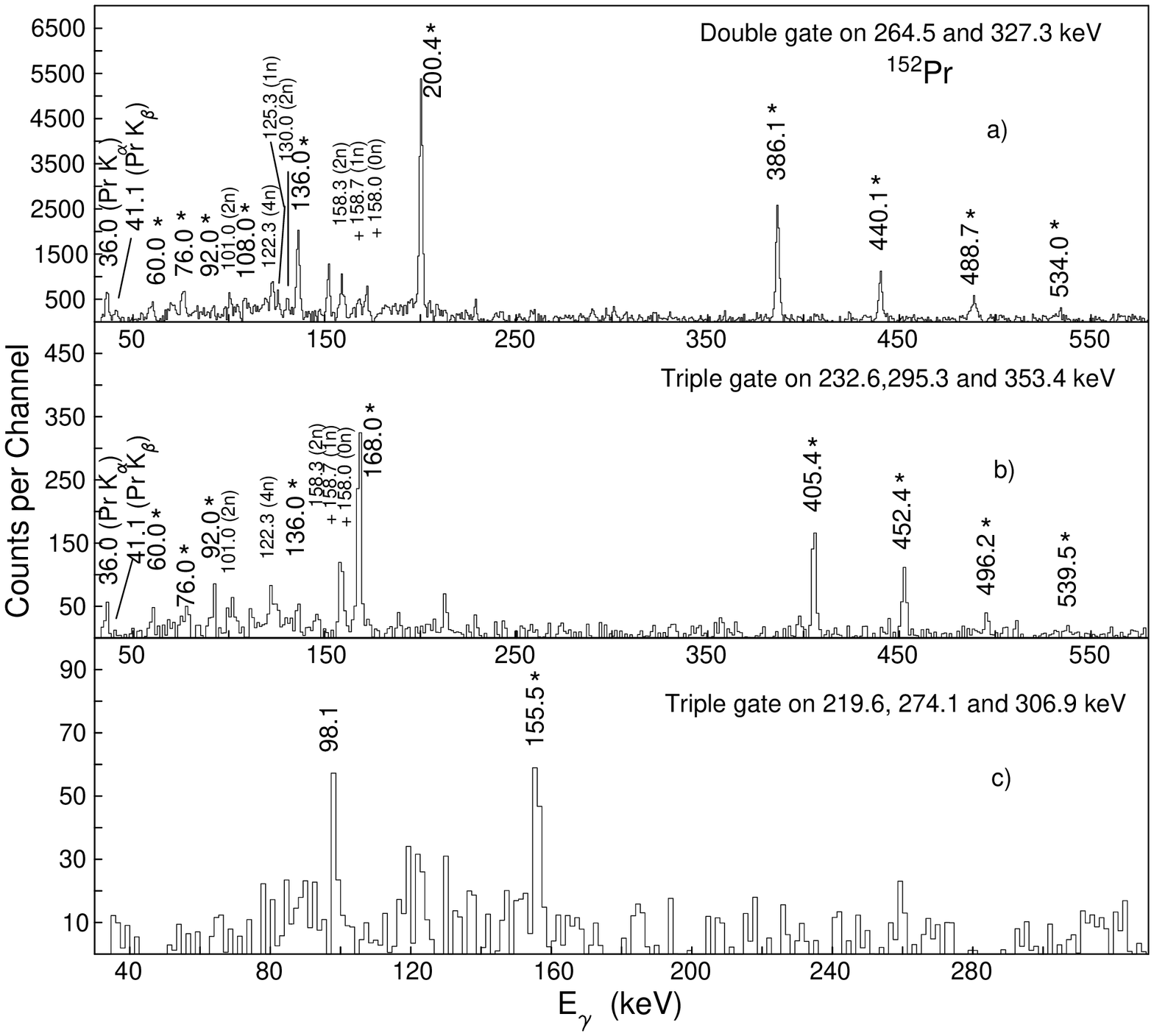}
 \caption{\label{152gate}Partial $\gamma$-ray  coincidence spectra (a)
   by  gating on 264.5  and 327.3~keV transitions,  (b) by  gating on
   232.6, 295.3 and 353.4~keV transitions, and (c) by gating on 219.6,
   274.1 and  306.9~keV transitions  in $^{152}$Pr from  $^{252}$Cf SF
   data. In  the spectrum, transitions belonging to  Y fission partner
   are indicated  with neutron evaporation  numbers, specifically, 2n,
   1n, 0n correspond to $^{98,99,100}$Y respectively. The * indicates new transitions.}
\end{figure*}
The    new    level    scheme    for   $^{152}$Pr    is    shown    in
Fig.~\ref{152level}. All levels  and transitions are newly established
in the  present work  except for the  114.8 and  98.1~keV transitions,
which  were  identified   in  $^{152}$Ce  $\beta$-  decay  \cite{Yam}.
Fig.~\ref{152mass}  shows  the $^{152}$Pr  A-  and  Z- gated  'single'
$\gamma$  spectrum from  $^{238}$U  + $^{9}$Be  induced fission  data,
illustrating the evidence for the  mass assignment for these three new
bands.  All the  strong transitions in these three  bands can be seen.
As can be  seen from the figure the 221.3,  221.9, 358.4 and 365.3~keV
transitions are not  seen in A = 152  spectrum (Fig.~\ref{152mass}) as
compared to  A = 151 gated  spectra (Figs.~\ref{151mass}(a),(b)). This
confirms that 142.3-221.9-296.4 ($^{151}$Pr) and 142.3-221.3-292.0~keV
cascades  do  not  belong  to  $^{152}$Pr as  previously  reported  in
Ref.~\cite{Liu} but  in $^{151}$Pr and  the latter does not  belong to
$^{153}$Pr  as  recently  reported\cite{Mal}.   The  position  of  the
162.3~keV transition in $^{151}$Pr (100 relative intensity) is labeled
in  Fig.~\ref{152mass}. The absence  of the  162.3~keV peak  (the weak
peak seen in the valley is  at 160.5 keV) indicates that bands (1) and
(2)  assigned to  $^{151}$Pr belong  there  and not  to $^{152}$Pr  as
recently  reported  in  Ref.~\cite{Mal}.   These  facts  give  further
evidence for the  mass assignment for the four  bands in $^{151}$Pr in
the present  work.  Further analysis  of the mass assignments  will be
reported  in the  $^{153}$Pr.  The  relatively strong  90 and  188 keV
peaks in Fig.~\ref{152mass} are not  seen in any of the SF coincidence
data connected  to bands (1)  and (2) in Fig.~\ref{152level}.   The SF
coincidence data do indicate a 188-279-322-385 keV cascade which could
form another band in $^{152}$Pr.

Fig.~\ref{152gate}(a) shows a coincidence spectrum from the $^{252}$Cf
SF  data by  double gating  on 264.5  and 327.3~keV  transitions.  The
136.0, 200.4,  386.1, 440.1, 488.7  and 534.0~keV transitions  in band
(1) and  60.0, 76.0, 92.0, 108.4~keV linking  transitions between band
(1)  and  band  (2)   can  be  seen.   Fig.~\ref{152gate}(b)  shows  a
coincidence spectrum from  the $^{252}$Cf SF data by  triple gating on
232.6,  295.3 and  353.4~keV transitions,  where 168.0,  405.4, 452.4,
496.2 and 539.5~keV  transitions in band (2) and  60.0, 76.0, and 92.0
linking  transitions  between band  (1)  and  band  (2) can  be  seen.
Fig.~\ref{152gate}(c) shows a  coincidence spectrum from $^{252}$Cf SF
data by triple gating on 219.6, 274.1 and 306.9~keV transitions, where
98.1 and 155.5~keV transitions in  band (3) can be seen. The 114.8~keV
transition is  not observed in  the current work since  the electronic
coincidence time window for the $^{252}$Cf fission experiment is about
1 $\mu$s,  which is relatively smaller  than the 4.1  $\mu$s life time
(\cite{Yam})  of the 114.8~keV  level.  For  prompt-$\gamma$-ray data,
the  present GANIL experimental setup  is  sensitive  only  to states  with
lifetimes  shorter   than  $\sim  2$~ns.   The   result  for  $\gamma$
transition intensities are also given in the level scheme.

\subsection{$^{153}$Pr}

\begin{figure}
 \includegraphics[width=0.8\columnwidth]{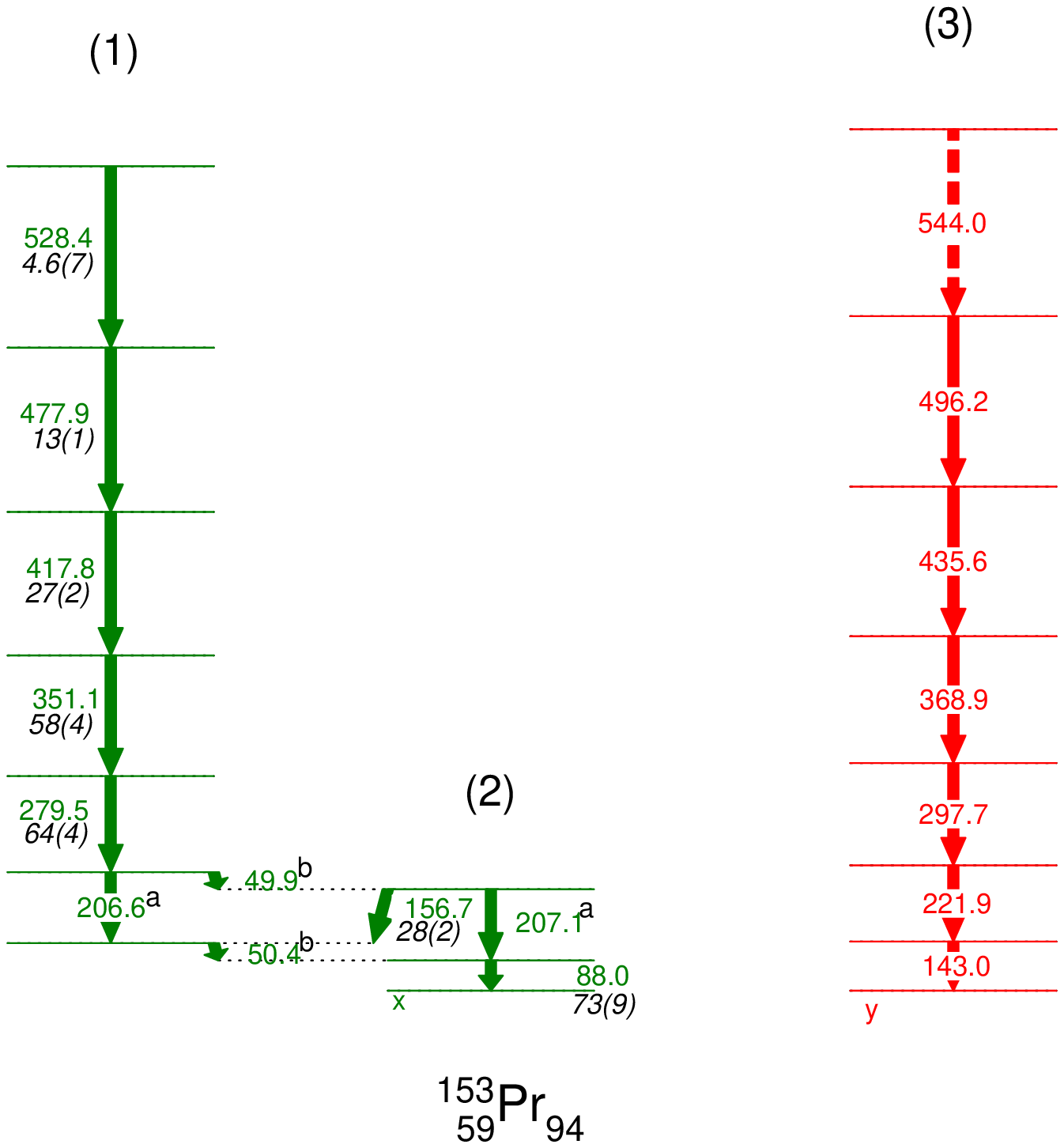}
 \caption{\label{153level}The  level  scheme   of  $^{153}$Pr  in  the
   present work.  Transitions and  levels previously reported by Hwang
   {\it  et  al.}\cite{Hwa10} are  labeled  in  green,  bands (1)  and
   (2).  New   ones  in  band  (3)   are  labeled  in   red.   a:  The
   206.6+207.1~keV transitions have the  relative intensity of 100. b:
   The  49.9+50.4~keV  transition   have  the  relative  intensity  of
   $<$127.  The new transitions  in red  were too  close in  energy to
   transitions in $^{151}$Pr to measure intensities.}
\end{figure}

\begin{figure}
\centering\includegraphics[width=\columnwidth]{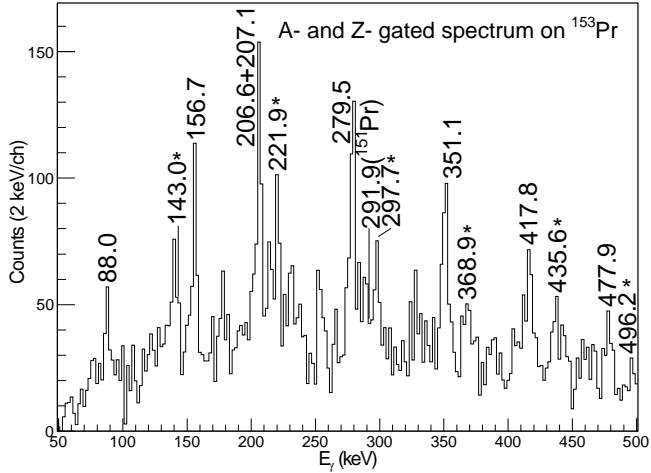}
\caption{\label{153mass}Partial $^{153}$Pr mass- and Z- gated 'single'
  $\gamma$-ray  spectra  obtained from  $^{238}$U  + $^{9}$Be  induced
  fission  data.   The  position  of  the 291.9  keV  transition  from
  $^{151}$Pr is also indicated  to illustrate its non-observation (see
  text for details). The * indicates new transitions. }
\end{figure}

The new  level scheme for $^{153}$Pr is  shown in Fig.~\ref{153level}.
Band (1) in Fig.~\ref{153level}  was assigned previously to $^{153}$Pr
in Ref.~\cite{Hwa10},  but was assigned, recently,  to $^{154}$Pr from
the SF  work of $^{248}$Cm  and $^{252}$Cf \cite{Mal}. In  the earlier
work,  the   relative  yield  ratios   of  partner  Y   isotopes  were
measured~\cite{Hwa10}.   The  $^{153}$Pr  mass  and  Z  gated  spectra
obtained  from  $^{238}$U +  $^{9}$Be  induced  fission  are shown  in
Fig.~\ref{153mass}.   The previously reported  $\gamma$-transitions of
88.0, 156.7,  206.6, 279.5, 351.1, 417.8 and  477.9~keV transitions in
$^{153}$Pr  \cite{Hwa10}  are confirmed  from  the  mass  and Z  gated
spectrum shown in Fig.~\ref{153mass}.   The 142.3, 221.3 and 292.2~keV
cascade previously  assigned to $^{153}$Pr by Malkiewicz  {\it et al.}
\cite{Mal}  is not  confirmed  by the  mass  and Z  gated spectrum  in
Fig.~\ref{153mass}. The position of  the 292.2~keV transition is shown
in Fig.~\ref{153mass}  to illustrate its  non-observation.  Instead, a
new cascade with 143.1, 221.9  and 297.7~keV transition can be seen in
the  spectrum.  Further  analysis about  the mass  assignment  will be
reported in the discussion part.

\begin{figure}
\centering\includegraphics[width=\columnwidth]{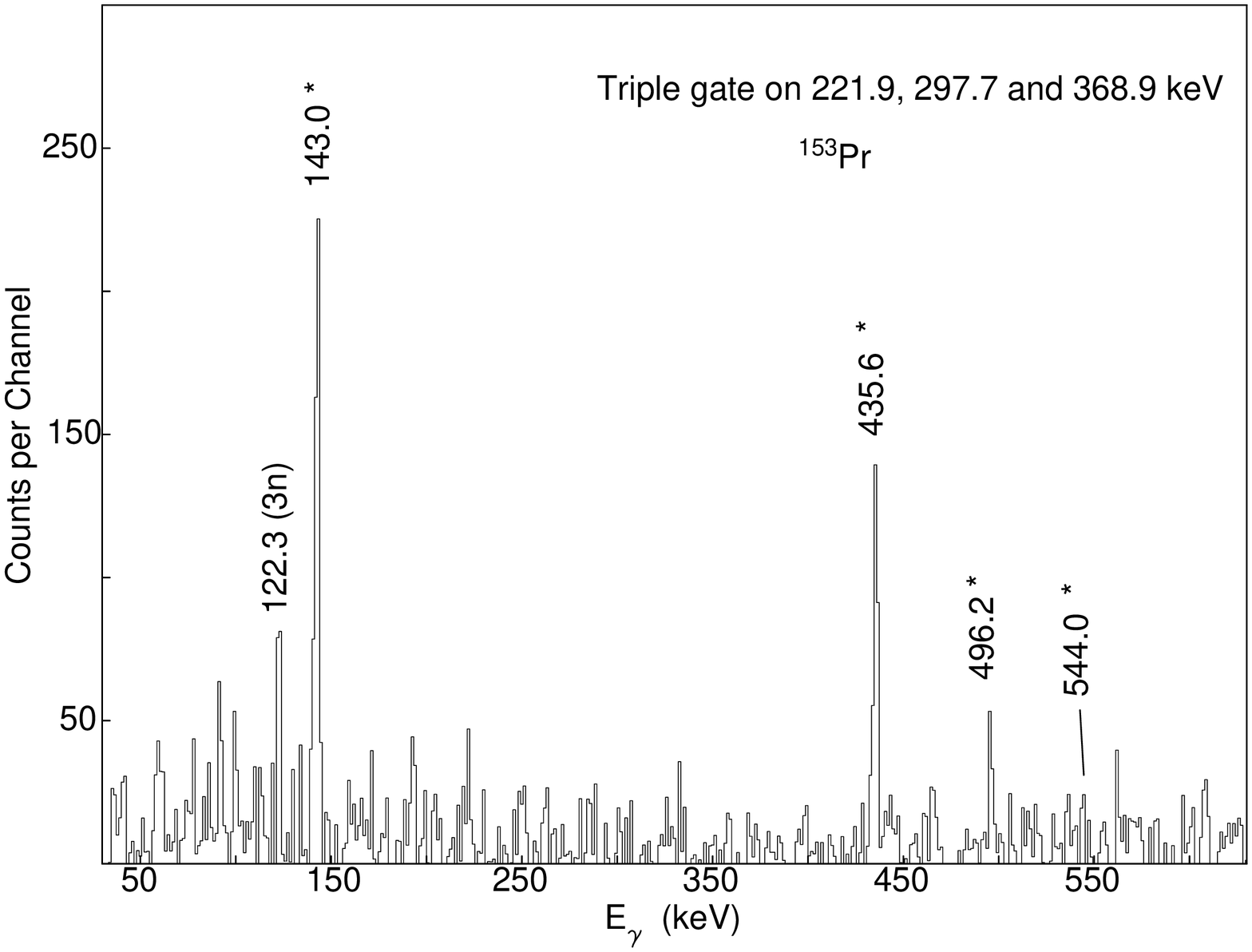}
\caption{\label{153gate}Partial  $\gamma$-ray coincidence  spectrum by
  triple  gating  on  221.9,   297.7  and  368.9~keV  transitions  in
  $^{153}$Pr from  $^{252}$Cf SF  data.  In the  spectrum, transitions
  belonging  to  the Y  fission  partner  are  indicated with  neutron
  evaporation numbers, specifically, 3n correspond to $^{96}$Y. The * indicates new transitions. }
\end{figure}

The previously assigned 51.7,  227.8 and 277.7~keV transitions are not
placed  in  Fig.~\ref{153level}  because  these  transitions  are  not
clearly seen  in the $^{252}$Cf  SF data.  Furthermore,  the 227.8~keV
transition  is not  observed in  the  206.6 and  969.1~keV (a  fission
partner $^{95}$Y transition\cite{Urb09}) double gate and the 206.6 and
417.8~keV double  gate. Thus, the  previously reported 13/2  level and
its  three   transitions  are  tentatively  removed   from  the  level
scheme. The  results for $\gamma$ transition intensities  are shown in
the   level  scheme.    Previous   internal  conversion   measurements
established  the 50.4~keV  transition as  E1 and  the 88.0~keV  one as
M1\cite{Hwa10}. The  value of the 50.4~keV  transition is questionable
because the  206.6/351.1~keV gate  used in Ref.~\cite{Hwa10}  might be
interfered by  the 207.1~keV transition. If  one approximately assumes
the  49.9  and 50.4~keV  transitions  have  the same  $\alpha$$_{exp}$
value,  the 2.3(7)  value  in Ref.~\cite{Hwa10}  should  be used,  not
3.2(9)  in   the  erratum  Ref.~\cite{Hwaer}   because  the  227.8~keV
transition is  not confirmed  in the current  work. Fig.~\ref{153gate}
shows a coincidence spectrum from  $^{252}$Cf SF data by gating on the
221.9,  297.7 and  368.9~keV transitions.   In this  gate,  four other
$\gamma$-ray transitions  in band  (3) can be  seen. The  (A,~Z) gated
spectra were important in guiding the $\gamma$ ray coincidence spectra
analysis to identify new band (3).

\section{discussion}
\begin{table}
\centering\caption{\label{table:1}Comparison  of  the  E2  transitions
  energies  in  bands  (1),  (2) and  (3)  in  $^{147,149,151,153}$Pr  and  the
  transition energy difference.}  \renewcommand{\arraystretch}{1.3}

\resizebox{\columnwidth}{!}{\begin{tabular}{cccc|ccc|ccc}
\hline\hline
\multicolumn{4}{c|}{band 1 E2 (keV)} & \multicolumn{3}{c|}{band 2 E2 (keV)} & \multicolumn{3}{c}{band 3 E2 (keV)} \\
\hline
$^{147}$Pr & $^{149}$Pr & $^{151}$Pr & $^{153}$Pr & $^{149}$Pr & $^{151}$Pr & $^{153}$Pr & $^{149}$Pr & $^{151}$Pr & $^{153}$Pr \\
\hline
& 103 & & & & & & & 142 & 143\\
256 & 220 & 216 & 207 & 232 & 215 & 207 & & 222 & 222\\
424 & 330 & 292 & 280 & 345 & 299 & & 279 & 296 & 298\\
536 & 416 & 363 & 351 & 437 & 377 & & 372 & 365 & 369\\
608 & 480 & 427 & 418 & 506 & 445 & & 437 & 429 & 436\\
660 & 523 & 483 & 478 & 536 & & & 491 & 487 & 496\\
& 535 & 532 & 528 & 492 & & & 527 & 541 & 544\\
& 520 & 575 & & & & & & 584 &\\
& 539 & & & & & & & &\\
\hline\hline
\multicolumn{4}{c|}{band 1 $\Delta$E2 (keV)} & \multicolumn{3}{c|}{band 2 $\Delta$E2 (keV)} & \multicolumn{3}{c}{band 3 $\Delta$E2 (keV)} \\
& 117 & & & & & & & 80 & 79\\
168 & 110 & 76 & 73 & 113 & 84 & & & 74 & 76\\
112 & 86 & 71 & 71 & 92 & 78 & & 93 & 69 & 71\\
72 & 64 & 64 & 67 & 69 & 68 & & 65 & 64 & 67\\
52 & 43 & 56 & 60 & 30 & & & 54 & 58 & 60\\
& 12 & 49 & 50 & -44 & & & 36 & 54 & 48\\
& -15 & 43 & & & & & & 43 & \\
& 19 & & & & & & & & \\
\hline\hline
\end{tabular}}

\end{table}
Mantica  \emph{et   al.}   showed  that   the  quadrupole  deformation
increases  gradually from  $^{145}$Pr to  $^{149}$Pr  \cite{Man}.  The
current   work  implies   $^{151,153}$Pr  have   similar   but  larger
deformation than  $^{149}$Pr based on  the decreasing E2  energies and
$\Delta$E2 values shown for their bands (1) in Table~\ref{table:1}.  A
comparison  of  the proposed  E2  transition  energies and  transition
energy spacing in bands (1),  (2) and (3) in $^{147,149,151,153}$Pr is
shown  in  Table~\ref{table:1}.   The  similarity  of  the  transition
energies indicates similar structures  of the nuclei.  Moreover, bands
(1) and (3) in $^{151,153}$Pr  are almost identical both in transition
energy (less  than 10~keV) and intensity  up to very  high spins. Such
kind   of    almost   identical   bands   were    also   observed   in
$^{152,154,156}$Nd    \cite{Smi},    $^{153,155,157}$Pm    \cite{Ran},
$^{156,158,160}$Sm    \cite{Zhu},    $^{155,157,159}$Eu    \cite{Bur},
$^{160,162,164}$Gd  \cite{Jon}  respectively.   This  identical  bands
phenomenon occurs  just after the  phase transition from  spherical to
large  deformed  shape   as  N  increases  from  88   to  90  in  this
region. $\Delta$E2 energy shrinks more in $^{149}$Pr and bending takes
place at high  spin. In contrast, the $\Delta$E2  energies are similar
and  do  not shrink  as  much  in  $^{151,153}$Pr, which  indicates  a
relatively  rigid  rotor  in  these  nuclei.  Note  that  the  E2  and
$\Delta$E2  energies in  bands  (1),  (2) and  (3)  in $^{150}$Pr  are
different to indicate that band (2) has a large deformation than bands
(1) and  (3).  Note the  E2 energies in  band (1) of  N=88 $^{147}$Pr,
Fig.~\ref{147level} are 36, 94, 120, 128, 137 keV, respectively larger
than in band 1 of N=90  $^{149}$Pr where the differences are only 4 to
53  keV  between  $^{149}$Pr   and  $^{151}$Pr  to  indicate  a  phase
transition   here  too.    An  internal   conversion   measurement  in
Ref.~\cite{Hwa10}  implied E1 interband  transitions between  band (1)
and band (2) in $^{151}$Pr. Therefore, there could be a small octupole
deformation  or  correlation  in  $^{151}$Pr  and  the  similar  cross
transitions  between  bands  (1)  and  (2) likewise  suggest  this  in
$^{149}$Pr.   QPRM  calculations   for  $^{149,151,153}$Pr  have  been
discussed earlier in Refs.~\cite{Rza,Mal}.

Pr nuclei in this region present serious challenges in assigning bands
to particular odd-Z Nilsson orbitals.  There are \emph{K}=1/2 orbitals
of  both  parities and  large  $j$ values  near  the  Fermi energy.  The
odd-parity \emph{K}=1/2$^-$[550] has \emph{j}=11/2 and the even parity
\emph{K}=1/2$^+$   orbitals  are   a  mix   of   \emph{g}$_{7/2}$  and
\emph{d}$_{5/2}$.   The odd-parity \emph{K}=1/2  band will  have large
signature splitting,  and the even parity \emph{K}=1/2  bands can have
either sign  of signature splitting  or near cancellation  for certain
admixtures.

The 368.8, 546.3 and 672.3~keV  transitions in the new 3/2$^+$ band in
$^{145}$Pr are similar to the first three E2 transitions (397, 541 and
709~keV) in $^{144}$Ce.  Therefore, the  aligned angular  momentum of
this   new  band   relative   to  the   $^{144}$Ce   core  is   around
3.0$\sim$3.5$\hbar$.  Such comparison  indicates the  3/2$^+$  band in
$^{145}$Pr  is  possibly  originated  from  the  $\pi$\emph{g}$_{7/2}$
orbital.

As discussed in the previous  part, band (1) in $^{147}$Pr is proposed
to have a negative parity. The available orbital of negative parity in
this  region   is  $\pi$\emph{h}$_{11/2}$.   According   to  the  PTRM
calculations \cite{Man}, 1/2$^-$[550] band  was proposed to be the low
lying negative parity  one. Note that the negative  parity band (1) in
$^{149}$Pr in  the current work  was also proposed to  be 1/2$^-$[550]
from previous QPRM calculation \cite{Rza}.

The band  (2) in $^{149}$Pr does  not match any  configurations in the
previous QPRM  calculations which are mentioned  here for completeness
\cite{Rza}.  This band could form a s=i octupole band with band (1) if
they have the  opposite parity. Note that octupole  correlation is not
included   in   the   QPRM   calculation   in   Ref.~\cite{Rza}.   Our
potential-energy-surface calculation in the present work (more details in
Ref.~\cite{Naz,Xu00,Xu02}) shows medium octupole deformation
($\beta$$_3$=0.068) of 1/2$^-$[550] configuration (band (1)). Although
the previous  QPRM calculations predicted this  configuration does not
have 100$\%$  amplitude (91$\%$ in Ref.~\cite{Rza},  77$\%$ and 81$\%$
in Ref.~\cite{Gab}), the octupole deformation may change the signature
splitting of  the 1/2$^-$[550] band (1) in  $^{149}$Pr. Furthermore, a
small  double  backbending  occurs   in  band  (1)  of  $^{149}$Pr  at
$\hbar$$\omega$$\sim$0.27 MeV while a  little more distinct one occurs
at about  the same rotational frequency  in band (2).  As discussed in
Ref.~\cite{Hwa00},  cranked   shell  model  calculations  \cite{Zhu99}
suggest  that  this  backbending  at  0.27  MeV  originates  from  the
alignment of  a neutron  $i_{13/2}$ pair and  not a  proton $h_{11/2}$
pair.  Thus, band  (2) in $^{149}$Pr can also  be another signature of
band  (1).  If  bands  (1) and  (2)  in $^{149}$Pr  have the  opposite
parities,   the   average    B(E1)/B(E2)   ratio   would   be   around
0.05$\times$10$^{-6}$ fm$^{-2}$.  This  value lies between the average
B(E1)/B(E2)   ratios    of   the   octupole    bands   in   $^{148}$Ce
(0.82$\times$10$^{-6}$        \cite{Che})        and        $^{150}$Ce
(0.04$\times$10$^{-6}$ \cite{Zhu12})  and is smaller than  that of the
positive     branch     of     octupole    bands     in     $^{147}$La
(0.38$\times$10$^{-6}$). These data suggest the importance of octupole
correlations in $^{149}$Pr as found in $^{148}$Ce \cite{Jon}. Note the
predicted  center   of  octupole   deformation  is  that   Z=56,  N=88
\cite{Naza}, so $^{147}$La with Z=57 and N=90 and $^{148}$Ce with Z=58
and  N=90 are  close to  the  center and  have the  expected large  E1
strength.   Here $^{149}$Pr  with Z=59  and  N=90 is  more similar  to
$^{150}$Ce with Z=58, N=92, so  the E1 strength decreases as one moves
farther away from the center of Z=56 and N=88.

Spin and parity assignments of  the bands in $^{151}$Pr and $^{153}$Pr
are not placed in the  present work because the E2 transition energies
in these bands are quite similar. According to the QPRM calculation in
Ref.~\cite{Mal},  3/2$^-$[541],   1/2$^+$[420]  and  3/2$^+$[422]  are
proposed to  be the three low  lying states in  $^{151,153}$Pr. If one
assumes  the   bands  (1)  and   (2)  in  $^{151}$Pr  have   the  same
configuration and parity, then  only the theoretical prediction of the
3/2$^+$[422]  configuration  in   Ref.~\cite{Mal}  can  reproduce  the
signature splitting in bands (1)  and (2).  According to the statement
in  the previous  part, the  x~keV level  in $^{151}$Pr  could  be the
3/2$^-$  ground   state.   The  x+182.5~keV  level  in   band  (1)  in
Fig.~\ref{151level} would be 7/2$^+$.  Bands (3) and (4) in $^{151}$Pr
could be 1/2$^+$[420] or 3/2$^-$[541] but 1/2$^+$[420] is more likely,
because no  linking transitions are  observed between bands  (1,2) and
(3,4) in  the current work  and bands (3)  and (4) are more  likely to
decay  to  the  35.1~keV  isomer.    The  y~keV  level  could  be  the
1/2$^+$[420] state.  The configurations  of bands in $^{153}$Pr can be
assigned    according   to    the   level    scheme    similarity   to
$^{151}$Pr. However,  such assignments  cannot explain the  absence of
bands  (2)  and  (4)  in $^{153}$Pr.   Also, our potential-energy-surface
calculation reported here shows  octupole  deformation ($\beta$$_3$=0.043)  of  the
1/2$^+$[420] configuration.   If bands (1) and (2)  in $^{151}$Pr have
the opposite parity,  they can form an octupole  band.  In the current
work, band  (1) is the strongest  populated one. Thus,  similar to the
discussion in Ref.~\cite{Mal}, states  in this band are possibly yrast
and more likely to be the  favored branch of the 3/2$^-$[541]. In all,
similar to  the discussion  of $^{149}$Pr, spins  and parities  of the
bands in $^{151,153}$Pr still cannot be firmly assigned.

Potential-energy-surface    calculations   show
tendencies  toward  both octupole  deformation  and triaxiality.   The
octupole Y$_{3,0}$  deformation can give rise to  parity doubling, and
triaxiality can give so-called $\gamma$ bands. Further theoretical and
experimental work is needed.

\section{Conclusion}

In conclusion, new bands in $^{145,147-150}$Pr have been reported; the
previous  questionable  assignments   of  transitions  and  levels  in
$^{151,153}$Pr were  clarified by  A- and Z-  gated spectra  and yield
curves;  a  new  high  spin  level  scheme  for  $^{152}$Pr  has  been
established and confirmed by  mass-Z gated spectra. Spins and parities
of  the levels in  the new  band in  $^{149}$Pr have  been tentatively
assigned.     New     transitions    have    been     identified    in
$^{143-146}$Pr. More work  on both experiment and theory  is needed to
understand the nuclear structure of $^{147-153}$Pr well. The levels in
$^{149,151}$Pr  are  similar  to  octupole  structures.  The  bands  in
$^{151,153}$Pr are very similar in transition energies and $\Delta$E2,
which indicates  they are relatively  rigid rotors.  The  long multiple
high  spin bands  with increasing  neutron number  in the  Pr isotopes
provide interesting tests for  nuclear model calculations.  The unique
combination of  (A,~Z) identified in-beam $\gamma$-rays  and high fold
data from a  Cf source has opened new vista to  study the evolution of
nuclear  structure   as  functions   of  spin  and   isospin.  Further
improvements in the sensitivity  for fission fragment spectroscopy are
planned  using the  next generation  tracking detector  AGATA combined
with an improved VAMOS++ spectrometer at GANIL.

\begin{acknowledgments}
The  work  at Vanderbilt  University  and  Lawrence Berkeley  National
Laboratory are  supported by the  US Department of Energy  under Grant
No. DE-FG05-88ER40407 and Contract  No. DE-AC03-76SF00098. The work at
Tsinghua  University was  supported  by the  National Natural  Science
Foundation of  China under  Grant No. 11175095.  The work at  JINR was
supported  by   the  Russian  Foundation  for   Basic  Research  Grant
No.  08-02-00089 and  by the  INTAS Grant  No. 03-51-4496.  One  of us
(S.B.)  acknowledges   partial  financial  support   through  the  LIA
France-India agreement.  We would like to thank J.~Goupil, G.~Fremont,
L.~M\'{e}nager,  J.~Ropert,  C.~Spitaels,  and the  GANIL  accelerator
staff for their technical contributions.
\end{acknowledgments}

\end{document}